\documentclass[pre,twocolumn,aps,a4paper,superscriptaddress,showpacs,10pt,nofootinbib,floatfix,english]{revtex4-1}
\usepackage{graphicx}
\pdfoutput=1
\usepackage{babel}
\usepackage{dsfont}
\usepackage{textcomp}
\usepackage{graphicx}
\usepackage{flafter}
\usepackage{amsmath,amssymb}
\usepackage{bm}
\usepackage{pdfsync}
\usepackage{memhfixc}
\usepackage{natbib}

\begin{document}

\newcommand{\reff}[1]{(\ref{#1})}
\newcommand{\eref}[1]{Eq.\reff{#1}}
\newcommand{\erefs}[1]{Eqs.\reff{#1}}
\newcommand{\Eref}[1]{Equation \reff{#1}}
\newcommand{\sref}[1]{Sec. \ref{#1}}
\newcommand{\fref}[1]{Fig.\ref{#1}}

\newcommand{\psiD}{\psi_{\mathrm{D}}}
\newcommand{\psiDo}{\psi_{\mathrm{D0}}}
\newcommand{\Mstar}{M_{\mathrm{*}}}
\newcommand{\Md}{M_{\mathrm{d}}}
\newcommand{\Msun}{M_{\mathrm{\odot}}}
\newcommand{\dMd}{\dot{M}_{\mathrm{d}}}
\newcommand{\muo}{\mu_{0}}
\newcommand{\vB}{\vec{B}}
\newcommand{\vJ}{\vec{J}}
\newcommand{\omegaK}{\omega_{\mathrm{K}}}
\newcommand{\omegaKo}{\omega_{\mathrm{K0}}}
\newcommand{\mD}[1]{{\mathcal{D}}[#1]}
\newcommand{\vA}{v_{\mathrm{A}}}
\newcommand{\vs}{v_{\mathrm{s}}}
\newcommand{\vshq}{\hat{v}_{\mathrm{s}}^2}

\newcommand{\vt}{v_{\mathrm{t}}}
\newcommand{\me}{m_{\mathrm{e}}}
\newcommand{\mi}{m_{\mathrm{i}}}
\renewcommand{\ne}{n_{\mathrm{e}}}
\newcommand{\nupe}{\nu_{\mathrm{pe}}}
\newcommand{\nuee}{\nu_{\mathrm{ee}}}
\newcommand{\rs}{R_{\mathrm{S}}}
\newcommand{\Rstar}{R_{\mathrm{*}}}
\newcommand{\Ls}{L_{\mathrm{s}}}
\renewcommand{\Lsh}{\hat{L}_{\mathrm{s}}}
\newcommand{\const}{\mathrm{const}}
\newcommand{\Kb}{K_{\mathrm{B}}}
\newcommand{\km}{\mathrm{km}}
\newcommand{\keV}{\mathrm{keV}}
\newcommand{\erf}{\mathrm{erf}}
\newcommand{\tk}{\tilde{k}}
\newcommand{\tm}{\tilde{m}}
\newcommand{\visco}{\mathds{D}}

\title{Crystalline Structure of Accretion Disks: Features of the Global Model}

\author{Giovanni Montani}
\email{giovanni.montani@frascati.enea.it} 
\affiliation{ENEA - C.R. Frascati, U.T.Fus. (FUSMAG Lab),
              Via Enrico Fermi, 45 (00044), Frascati (Roma), Italy}
\affiliation{Dipartimento di Fisica Universit\`a di Roma ``Sapienza''
Piazzale Aldo Moro 5, 00185, Roma, Italy}
\affiliation{INFN - Istituto Nazionale di Fisica Nucleare, Sezione di Roma 1}

\author{Riccardo Benini}
\email{riccardo.benini@gmail.com}
\affiliation{Dipartimento di Fisica Universit\`a di Roma ``Sapienza''
Piazzale Aldo Moro 5, 00185, Roma, Italy}

\today

\begin{abstract}

In this paper, we develop the analysis of a two-dimensional magnetohydrodynamical configuration for an axially symmetric and
rotating plasma (embedded in a dipole like magnetic field), modeling the structure of a thin accretion disk
around a compact astrophysical object.
Our study investigates the global profile of
the disk plasma, in order to fix the conditions
for the existence of a crystalline morphology and ring sequence,
as outlined by the local analysis pursued in \cite{C05,CR06}.
In the linear regime, when the electromagnetic
back-reaction of the plasma is small enough,
we show the existence of an oscillating radial behavior
for the flux surface function which very
closely resembles the one outlined in the local model,
apart from a radial modulation of the amplitude.
In the opposite limit, corresponding to a dominant
back-reaction in the magnetic structure over the
field of central object, we can recognize the existence
of a ring-like decomposition of the disk,
according to the same modulation of the magnetic
flux surface, and a smoother radial decay of the disk density,
with respect to the linear case.
In this extreme non-linear regime, the global model
seems to predict a configuration very close to that of the local
analysis, but here the thermostatic pressure, crucial
for the equilibrium setting, is also radially modulated.

Among the conditions requested for the validity
of such a global model, the
confinement of the radial coordinate within a given value sensitive to the disk temperature and to the mass of the central objet, stands; however, this condition corresponds to dealing with a thin disk configuration.
\end{abstract}


\maketitle

\section{Introduction}

The understanding of the accretion process \cite{Bisno01} characterizing compact astrophysical objects is regarded as a central issue when describing the triggering of high energy phenomena, such as the  formation of jets.
 Indeed, it is commonly believed that the intense  electromagnetic emission in different bands from astrophysical sources, such as Gamma Ray Bursts (see for example \cite{ReviewPiran}) or Active Galactic Nuclei (see for example \cite{LibroAGN}), should be the final result of the proper instabilities of the accretion mechanism. Theoretical attempts have been made over the years to explain the formation of jets from the accretion profile of compact and massive astrophysical systems (see, among others, \cite{L96} and \cite{Sp08}).

The accretion mechanism in astrophysical systems is, however, far from being settled, because many aspects of the angular momentum transport within the gravitating plasma are still open to scientific debate.
The stellar disk configurations, on which the present analysis is focused, are described by a well-established paradigm, which, for the case of a thin disk profile, has the character of a standard model \cite{Bisno01}. This scenario satisfactorily reproduces the phenomenology fixed by the observational data, but at the price of postulating a viscoresistive magnetohydrodynamic (MHD) approach, which is not supported by the microscopic features of the disk plasma. The possibility to reconcile such an effective description of the non ideal nature of the plasma with the microscopic estimations of the viscosity and resistivity coefficients is individuated in the appearance of strong turbulence behaviors inside the disk. The theoretical framework for describing the onset of the plasma instabilities is offered by the so-called magnetorotational instability, introduced in \cite{V59, Cha60} (see also \cite{Bal91, B98}).

Even if the standard model for the thin accretion disks is rather successful and widely accepted, it cannot escape the following two relevant criticisms.
 (i) The $z$-dependence of the configuration is somewhat frozen out from the equilibrium by an average procedure along the vertical direction of both the radial and azimuthal equations, as originally proposed in \cite{S73,Shakura:1973p110}. 
(ii) The experience acquired in plasma morphology from the laboratory activities, does not confirm the direct relation between the existence of a turbulent regime and the applicability of the viscoresistive MHD scenario \cite{C94}.

In \cite{Ogilvie97} it is shown how the configuration of a thin disk, treated in a two-dimensional ideal MHD scheme, reveals very different features with respect to the effective one-dimensional standard model. In fact, a new equilibrium profile, corresponding to a strong magnetization of the disk, is determined by virtue of the confinement induced by the Lorentz force. The main significant property of this regime is the non-Keplerian character of the angular velocity of the disk.

A more specific criticism to the standard model is pursued in \cite{C05,CR06}, where a fundamental description of the rotating disk plasma is addressed by postulating the ideal two-dimensional MHD and arguing that the accretion mechanism can be driven by intermittent ballooning instabilities which push the plasma toward the central object via its porosity near the $X$-points of the magnetic configuration \cite{cproceeding}.
Until now, the main achievement of this new point of view,  is the demonstration that the local configuration around a fixed value of the radial coordinate, outlines the decomposition of the disk into a sequence of rings, corresponding to adjacent and opposite current density filaments. This issue takes place only in the limit of a very  strong electromagnetic back-reaction of the plasma (its existence in the limit of large values of the plasma $\beta$ parameter has been investigated in \cite{lattanzimontani10}), but a periodic (crystalline) structure of the magnetic surface functions already emerges  in the linear regime.

In the present analysis, we fix the conditions necessary to recover the crystalline and ring like
profile within a \textit{global} two-dimensional MHD model. In other words, we quantify the role played by the radial envelope in determining the amplitude scaling of the oscillating behavior that characterizes the magnetic configuration.

In the linear case, we are able to sketch a clear picture for the emergence of the crystalline profile from the disk configuration, establishing that the amplitude of the perturbed magnetic surface scales as $\sqrt{r}$, while the mass density on the equatorial plane decays as $r^{-3}$ ($r$ being the radial coordinate). In order to obtain this scheme from the generic axially symmetric MHD scenario, we need to make some significant assumptions, which, however, can be mainly summarized by the thin nature of the disk, the short scale of the perturbations in the plasma, and by the confinement of the configuration well-inside a given radial region, determined by the model parameters (such as the mass of the central object and the disk temperature).

When the electromagnetic back-reaction within the disk plasma induces a dominant magnetic field (with respect to the one provided by the central object), we recover the same ring like decomposition described in the local model \cite{CR06}. The radial modulation of the amplitude of the oscillations of the magnetic surfaces is the same as in the linear case, but now the mass density of the disk decays slower along the
radial direction, i.e., like $r^{-1/2}$ instead of $r^{-3}$. Furthermore, in the non linear case, the pressure assumes a significant role in fixing the equilibrium, and its oscillation amplitude
is also modulated as $r^{-1}$, while retaining the same structure as in the local model.

The main achievement of the present analysis is to show that the local features outlined in \cite{C05}
for the steady configuration of a thin disk
have a global character too and therefore they may
concern fundamental astrophysical processes, taking
place within the disk plasma, such as the realization
of material jets \cite{CoppiHighEnergy,MontaniCarlevaro10}.
Furthermore, we provide a precise constraint
about the internal radial region in which
the crystal profile of the magnetic field
can take place, identifying a parameter discriminating for specific astrophysical sources.

The paper is organized as follows. In \sref{sec:necessity}, we give the motivation for a reformulation of the accretion problem toward a more ideal nature of the gravitating plasma. In \sref{sec:fundeq}, we review the fundamental equations of axially-symmetric two-dimensional MHD describing an accretion disk embedded in the gravitational and magnetic fields of a central object; furthermore we fix the relation between the angular frequency of the plasma and the flux function. In \sref{sec:persc}, we develop the perturbation scheme for the considered problem, expanding the configuration equations up to the first order in agreement with the assumption of dealing with small-scale perturbations, and fix the conditions needed to recover the crystalline structure in the global case. In \sref{sec:lincas}, we assess the linear case when the back-reaction of the plasma is much smaller than the dipole-like field of the central object, and, in \sref{sec:phenomenological}, we discuss the phenomenological implications of the constructed model in order to outline the main features of the global plasma configuration. Then, in \sref{sec:extreme}, we analyze the opposite case, i.e., when the back-reaction of the plasma cannot be considered as a small perturbation but it is the dominant contribution in fixing the global profile.
 Concluding remarks follow in \sref{sec:conrem}.

\section{Necessity for a Reformulation of the Accretion Picture}   
\label{sec:necessity}

In the \emph{Standard Model}
for the stellar accretion \cite{S73,Bisno01},
the configuration of an axisymmetric thin disk
is determined by the fluidodynamical equilibrium
which takes place in the gravitational field
of the compact accreting object (having mass $\Mstar$),
but performing an average procedure along the vertical direction.

The radial equilibrium states that the angular frequency
of the disk takes the Keplerian profile
$\omega(r)=\omegaK=\sqrt{G\Mstar/r^3}$. Significant
deviations from such a behavior are expected only
in advective dominated regimes.

The vertical equilibrium in the isothermal disk of temperature $T$,
fixes the exponential decay of the mass density $\rho$ over the
equatorial plane value $\rho_0(r)$, i.e.,
$\rho /\rho _0 \equiv D(z^2) =
\exp \{ -z^2/H^2\}$, where we introduced the typical length 
$H^2 = 2\vs^2/\omegaK^2$ estimating the half-depth
of the disk ($\vs$ is the sound velocity
on the equatorial plane, namely
$\vs^2=2\Kb T/\mi$, with $\Kb$ the
Boltzmann constant and $\mi$ the ion mass).
The azimuthal equilibrium describes the angular
momentum transport across the disk, by virtue of
a turbulent viscosity coefficient $\visco$, such that
\begin{equation}
\dMd(L - L_\mathrm{d}) = 3\pi \visco \omegaK r^2
\, ,
\label{amt}
\end{equation}
where $L$ is the angular momentum per unit mass,
$L_\mathrm{d}$ is a fixed value and
$\dMd=-2\pi r\Sigma v_r$ is the mass accretion
rate, associated with the radial equatorial velocity $v_r<0$ and  the
surface mass density $\Sigma \equiv \int_{-H}^{H}\rho dz$.
Finally, the continuity equation implies that
$\dMd = \const>0$ (a discussion of these equations can be found, for example, in \cite{Bisno01}).

\subsection{The viscoresistive puzzle}  \label{sec:puzzle}

The microscopic plasma structure accounts for
a too small viscosity coefficient $\visco$ arising in the disk
to explain the accretion rates observed
in some astrophysical systems, such as X-ray binaries.
In fact, the observed accretion rates, 
evaluated by the increasing disk luminosity
${\dot{L}}_\mathrm{d}\sim G\Mstar\dMd/\Rstar$, 
require large values of $\visco$, that in \cite{S73}
were associated with a postulated turbulent behavior
of the disk fluid. Since by definition
$L=\omega r^2$, we can infer
$\visco = 2\Sigma \vt H/3$, $\vt$ being a turbulence
velocity, expressible as $\vt=\alpha \vs$,
where $\alpha$ is a free parameter.
The fundamental question is 
whether the axisymmetric disk is linearly stable with
respect to small perturbations that preserve its symmetry.
A solution to this problem comes from the presence
of a non-vanishing magnetic field, which makes the non-linear interaction of very
small disturbances of the equilibrium possible.
Such a MHD instability, commonly known as
\textit{Magneto-Rotational Instability} (MRI), is triggered by the radial gradient of the disk
angular velocity and has been fixed by the Velikov analysis of 1959 \cite{V59,Cha60} (for a review on the topic, see \cite{B98}).
In \cite{Coppi:2008p98} it is
argued that the MRI is
strongly suppressed when the disk is
thin enough and therefore its efficiency in generating
turbulence is ruled out in favor of a \textit{thermorotational}
instability, in which the vertical gradient of the
temperature plays a crucial role.

Indeed, the powerful scenario emerging from the MRI relies on the presence of an even small, embedded magnetic field. However, the existence of an unstable mode of wavenumber $k$ is subjected to the condition
\begin{equation}
k^2 \vA^2 + \frac{d \omega ^2}{d\ln r} \le 0\,,
\quad \quad \omega (r) \simeq \omegaK =
\sqrt{\frac{G\Mstar}{r^3}}
\, ,
\label{MRIcond}
\end{equation}
where $\vA$ denotes the background Alfv\'en velocity. This simple condition can be easily recovered when the space dependence of the perturbation is parallel to the direction of magnetic field, here assumed without a significant loss of generality, along the $z$-axis. Condition \reff{MRIcond} can be recast in a more significant shape, as far as we realize that the stability of the disk profile as a whole requires the extension of such inequality to scale of the disk depth. A global disk stability is, in fact, ensured only if the Alfv\'en term dominates even for the smallest available wavenumbers $k \simeq \pi \omegaK/\sqrt{2}\vs$. Hence, observing that $d\omegaK^2/d\ln r = -3\omegaK^2$, we finally get
\begin{equation}
\vA^2 \le \frac{6}{\pi ^2} \vs^2
\, .
\label{MRIcond2}
\end{equation} 
As far as the magnetic field being sufficiently small and the disk thick enough, the emergence of an unstable mode is guaranteed. However, if the magnetic field of the central object is important, so that the Alfv\'en velocity is not too small and at the same time, the temperature and angular rotation of the disk are able to produce a sufficiently thin profile (for which the sound speed is constrained by a small upper bound $\vs /(\omegaK r) \sim H/r\ll 1$), behaviors violating condition \reff{MRIcond2} must be taken into account. In other words, we cannot exclude the existence of a class of thin disks for which the plasma admits a parameter $\beta$ less than a few units. Indeed, in the isothermal case we have
\begin{equation}
\beta\equiv 2\displaystyle\frac{\vs^2}{\vA^2} \leq \displaystyle\frac{\pi^2}{3}\,.
\end{equation}

For such relatively cold and strongly magnetized plasma disks, the MRI is suppressed and we need a new type of instability to account for the turbulent scenario required to deal with significant dissipative effects. Interesting features in favor of new instability perspectives are discussed in \cite{Coppi:2008p98,2000ApJ...529..978B,Coppi:2001p102}.

The presence of an intense magnetic field (as required by a pulsar-like central object) however, poses a new puzzle regarding the value of the non-zero resistivity coefficient $\eta$ of the ideal plasma. In fact, the equation of the electron force balance reads as
\begin{equation}\label{eq:3}
\vec{E} + \frac{\vec{v}}{c}\wedge \vec{B} = \eta \vec{J}\,,
\end{equation} 
with $\vec{E}$ and $\vec{B}$ denoting the electric and magnetic field, respectively, while $\vec{v}$ is the plasma velocity and $\vec{J}$ is the current density.
Since the axial symmetry requires  $E_{\phi }\equiv 0$, the azimuthal component  stands as
\begin{equation}\label{pippoplutopaperino}
v_z B_r - v_r B_z = c\eta J_\phi\,.
\end{equation}

In the one-dimensional model, obtained for the thin
disk by averaging along the $z$-axis, and taking
into account both the relation between $v_r$ and the
constant accretion rate $\dMd$, as well as the dipole-like
morphology of the magnetic field, \eref{pippoplutopaperino} is given by
\begin{equation}
\frac{\dMd\muo}{2\pi \Sigma (r)r^4} =
\eta cJ_{\phi}
\, , 
\label{releta}
\end{equation}
$\muo $ being a parameter modulating the intensity of
the dipole field. Observing that the superficial density is given by \cite{Bisno01}
\begin{equation}
\Sigma(r) = \left(\frac{2\pi \vs^2}{G\Mstar}\right)^{1/2}\rho_0(r) r^{3/2}\, ,
\label{supden}
\end{equation}
the azimuthal current density $J_{\phi}$ takes the radial behavior
\begin{equation}
J_{\phi} = \frac{\dMd\muo}{\eta c \vs}
\sqrt{\frac{G\Mstar}{(2\pi )^3}}  
\frac{1}{\rho _0(r)r^{11/2}}\, .
\label{releta2}
\end{equation}

\begin{figure}[ht]
   \centering
         \includegraphics[width=\columnwidth]{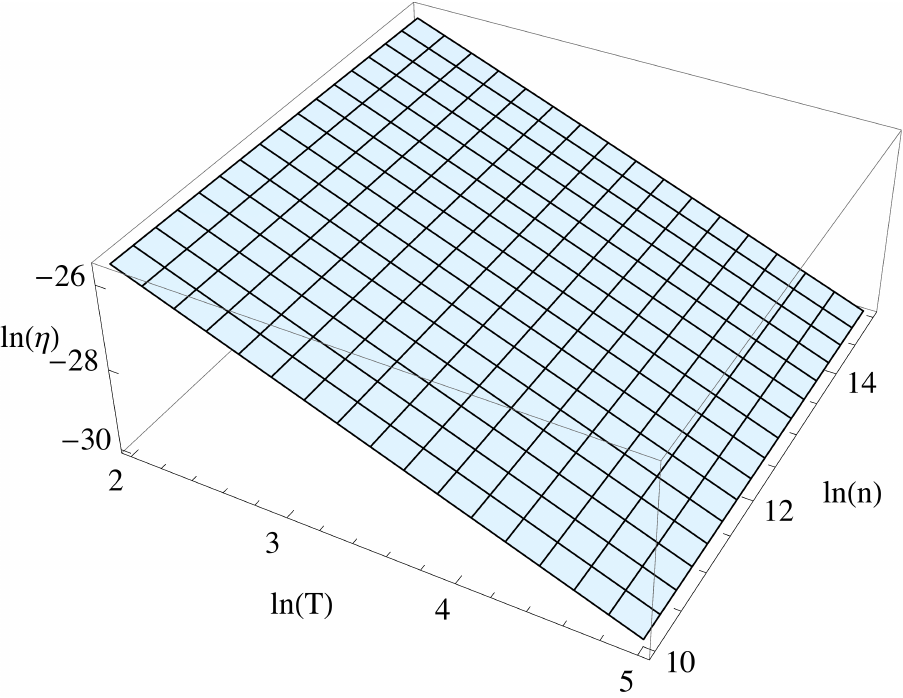} 
   \caption{Resistivity coefficient $\eta$  as a
function of the electron number density $\ne$ ($cm^{-3}$) and of the temperature of the
plasma $T$ (eV).}
   \label{fig:resistivita}
\end{figure}

The puzzle consists of the fact that the
strong dipole magnetic field is associated with
a zero density current (being a vacuum solution), and
thus the toroidal current $J_{\phi}$ must be
rather small since its existence is due to
the plasma backreaction only. The key
quantity we have to focus on is therefore
the ratio $\dMd\muo /\eta$, which is
requested to be very small as well.
For an $X$-ray star, for which the parameters
$\dMd$ and $\muo$, as determined from the
observations, are significantly high, the smallness
of the toroidal current $J_{\phi}$ can be reached only
for a sufficiently high value of the resistivity
coefficient. But this is clearly not the case,
as far as we microscopically estimate this
coefficient, accordingly to the expression (in Gauss CGS units, i.e., in seconds)
\begin{equation}\label{resistivitasec}
\eta = \displaystyle\frac{\me \nupe}{\ne e^2}\,,
\end{equation}
where $\me$ is the electron mass, $\nupe$ is the proton-electron collision frequency, $\ne$ is the electron number density, and $e$ is the electron charge. The proton-electron collision frequency $\nupe$ can be well approximated through the electron-electron collision frequency $\nuee$, given by\footnote{see the \textit{NRL Plasma Formulary} (Naval Research Laboratory, Washington, D.C., 2009).}
\begin{equation}\label{nuee}
\nuee=2.91\times 10^{-6} \left(\displaystyle\frac{\ne}{1\, \mathrm{cm^{-3}}}\right) \left(\displaystyle\frac{T}{1\, \mathrm{ eV}}\right)^{-\tfrac{3}{2}}\log(\Lambda) \mathrm{Hz}\,,
\end{equation}
where the Coulomb logarithm $\log(\Lambda)$ can be estimated via the formula
\begin{equation}\label{logaritmo}
\Lambda = 12 \pi N_\mathrm{D}\,,
\end{equation}
$N_\mathrm{D}$ being the Debye number of the plasma. In \fref{fig:resistivita}  this resistivity coefficient is plotted against the number density and the temperature of the plasma, showing the smallness of the corresponding assumed values.

The viscoresistive puzzle is therefore the combination
of two related facts: a magnetic field
is required by the MRI,
which allows the onset of turbulence, to generate the
viscosity term (without such dissipative effect no angular
momentum transport is permitted); this magnetic field  in turn requires
an anomalous resistivity to get a satisfactory balance
for the electron force along the azimuthal component.
We are led, thus, to postulate both a high viscosity
and resistivity coefficients, i.e., a magnetic Prandtl number
of order unity \cite{Sp08}. However,
no firm explanation exists for the possibility that
the turbulent behavior, resulting from the MRI, is able to generate significant dissipative effects in a thin disk, and, as discussed above, some plasma configurations can exist for which the MRI does not work at all.

\subsection{The crystalline structure: from the local toward a global model}

This apparent inconsistency of the
standard model for stellar accretion (at least for sufficiently magnetized astrophysical sources) and the possible failure of the MRI for a sufficiently thin disk, led Coppi in \cite{C05,CR06} to
reformulate the thin disk equilibrium in a pure
two-dimensional ideal MHD scenario. The point of view
proposed corresponds to
replacing the diffusive morphology of the magnetic field
in the plasma, resulting from the effective dissipative
processes, by rigid structures in the magnetic flux surfaces
and in the mass density profile, the so-called
\emph{crystalline structure}, with the associated
ring sequence decomposition of the disk.
In this scenario, the angular momentum transport is
not ensured by the turbulent viscosity effect, but
is argued to come from the porosity of the plasma
near the $X$-points of the magnetic configuration,
where the $z$-component of the $\vec{B}$ field vanishes. Indeed, near
such peculiar configurations, the radial velocity
can increase without any violation of the electron
force balance. In such a framework,
the plasma is pushed to infall by the action of intermittent
thermorotational unstable modes, similar to the
\emph{ballooning} modes observed in laboratory
experiments \cite{ballooning2,Ballooning1};
for a discussion of this inferred alternative scenario
for the disk accretion, see \cite{cproceeding}.

We also stress how the possibility to reconcile the crystalline
structure with the viscoresistive MHD scenario is discussed in \cite{2010GReGr.tmp..112M},
showing how the radial oscillation of the magnetic flux
surfaces can be recognized, for a local model, even in the presence of
 dissipative effects. Furthermore, \cite{2010PhRvE..82b5402M}  studied the ideal MHD
configuration of a thin disk near a fixed radius from the central object, but differently from the original works \cite{C05,CR06}, the
contribution of small poloidal currents and matter fluxes are retained in the problem.
The idea proposed concerns the possibility to balance the electron force equation by requiring the peculiar condition $v_zB_r - v_rB_z \simeq 0$ (to be intended as a net average prescription). In the framework of
a plasma decomposed in a ring sequence, the possibility to deal with strong values
of the vertical velocity, in correspondence to
suitable restrictions, is shown. The establishment of
such configurations seems to be of relevance in the
explanation of stellar wind profiles, as well as
in the determination of seeds for the jet formation.

All these results on the crystalline structure, however, are derived 
 within the scheme of a local model,
in which the plasma configuration is fixed around
a fiducial radius from the central object. 
The main goal of our analysis is the extension of the original local model to the global configuration of the disk. 

Indeed, the local analysis is developed around a fixed value of the radial coordinate in the disk, and therefore it is unable to clarify how the differentially rotating layers determine the radial behavior of the fundamental physical quantities, such as the pressure, the mass density and the magnetic flux surfaces. One of the main questions left open is whether the crystalline structure and the ring sequence arising in the disk, are compatible with a regular radial envelop, i.e., if these peculiar features can be reconciled with a physically meaningful behavior of the fundamental configuration variables. For instance, an important point to be addressed, and successfully handled in this paper, is the possibility to have a radial oscillation of the mass density (characterizing the ring formation) but in the framework of a global decay of the radial matter distribution as the outer regions of the disk are approached.

Having this scenario in mind, we formulate the global disk plasma equilibrium starting from those natural assumptions which are the expected generalization of what was proposed in \cite{C05,CR06}. As first step, we need to fix the explicit dependence that the plasma angular frequency outlines, according to the corotation theorem \cite{F37}, with respect to the background magnetic surface; the latter, in particular, are taken in the suitable form of a dipolar-like configuration. In the local model, this information is somehow hidden in a constant term and ultimately restated as the derivative of the angular frequency with respect to the radial variable evaluated at the fixed point. Here we fix the dependence $\omega = \omega(\psi)$ by determining it on the equatorial plane of the disk and then extending this relation to the whole plasma profile.

Another basic assumption at the ground of the global analysis presented below, is the request of dealing with a very small-scale backreaction of the plasma with respect to the magnetic field of the central object. In other words, we assume that the physical quantities (i.e., the internal currents, the mass density, pressure perturbations and the correction to the magnetic surfaces) have a very rapid radial variation, in agreement with the MHD. This request is equivalent to fixing a hierarchy in the radial gradient profile, and has the same impact on the plasma configuration as the smoothness of the plasma around the fixed radius, which is implicitly postulated in \cite{C05}.

These two requirements at the ground of the global model, i.e., the dipole like nature of the central object magnetic field and the very small scale of the plasma backreaction, appear as very natural choices for the physical characterization of a stellar disk, and their successful implementation to obtain the crystalline array of the plasma along the radial direction, must be accompanied only by the restriction to deal with a thin disk; such an assumption (already present in the local analysis)  is allowed by the many observations that such a class of disks is indeed present around compact and rapidly rotating astrophysical objects.

The real new contribution of what is discussed here, is the determination of the radial profile of the disk and the estimation of the microscopic nature (for an astrophysical setting) of the wavelength of the perturbations responsible for the ring decomposition of the disk. Such an analysis has no direct relation with \cite{2010GReGr.tmp..112M}, where the poloidal currents and matter fluxes are taken into account; however, the natural development of the present global model
would be to verify the existence of such peaks in the vertical velocity, once the radial envelope is fully accounted for. This scenario would allow the exact determination
of the disk regions where the stellar wind or the jets could take place.
Nonetheless, involving  poloidal currents
and matter fluxes in a global model would lead to non-trivial questions concerning
the validity of the corotation theorem, well beyond the
scope of the present approach. In fact, in the addressed toroidal
picture, the validity of such a theorem stands as well and we
are focusing our attention on the radial dependence of the crystalline structure, i.e., on how the magnetic flux surfaces and the mass density oscillate across the global radial profile of the disk.

\section{Fundamental equations}\label{sec:fundeq}

We now fix the fundamental configuration equations
for the axial symmetry of the disk, by considering
the central astrophysical object as a star of mass $\Mstar$, endowed with an intense magnetic field $\vec{B}$. The magnetic field can be taken in the form
\begin{equation}
\vB = -\frac{1}{r}\partial _z\psi \hat{e}_r +
\frac{1}{r}\partial _r\psi \hat{e}_z
\, , 
\label{vectorb}
\end{equation}
$\psi = \psi(r,z^2)$ being the magnetic flux function,
while the associated current density $\vJ$ remains fixed as
\begin{equation}
\vJ = - \frac{c}{4\pi }\left[ \partial _r\left(
\frac{1}{r}\partial _r\psi \right)
+ \frac{1}{r}\partial ^2_z\psi \right] \hat{e}_{\phi }
\, . 
\label{current}
\end{equation}
The magnetic field of the central object
is well described in the disk by a dipole-like
configuration, corresponding to the flux function
\begin{equation}
\psiD(r,z^2) = \frac{\muo r^2}
{\left( r^2 + z^2\right) ^{3/2}}\,,
\quad  \quad \mu _0 = \const\,.
\label{dipcurrent}
\end{equation}
Here, the constant value $\muo$ 
fixes the
dipole field amplitude and we stress that the
current density associated with this vacuum
configuration is characterized by the relation
$\vec{J}(\psi = \psiD) \equiv 0$.

The Newton potential $\chi$ describing the gravitational
field generated by the central object stands as
\begin{equation}
\chi(r,z^2) = \frac{G\Mstar}
{\sqrt{ r^2 + z^2}}
\, ,
\label{Newtpot}
\end{equation}
$G$ being the Newton constant.
We can also define the Keplerian angular velocity $\omegaK$ as
\begin{equation}
\omegaK^2 (r,z^2) = \frac{G\Mstar}
{\left(r^2 + z^2\right)^{3/2}} 
\, .
\label{angvel}
\end{equation}
We observe that on the equatorial plane $z = 0$,
the following relations hold
\begin{equation}
\left.
\begin{aligned}
\omegaK^2 (r,0) \equiv \omegaKo^2 = \frac{G\Mstar}{r^3}\\
\psiD = \frac{\muo}{r}
\end{aligned}
\right\}
\hspace{2mm}\Rightarrow \,\hspace{2mm}
\omegaKo^{2} = \frac{G\Mstar}{\muo^3}\psiD^3
\, .
\label{angvelvp}
\end{equation}
These relations are of interest because of the
thin nature of the disk.
Furthermore the corotation theorem \cite{F37} states that
the plasma angular velocity must be a function
of the magnetic surface only, i.e.,
$\omega = \omega(\psi)$, and it is natural to
postulate that the relation
\begin{equation}
\omega^2 = \frac{G\Mstar}{\muo^3}\psi^3\,,
\end{equation}
holding on the
equatorial plane, is valid everywhere in the disk. We note that, in agreement with
\eref{angvel}, the disk, embedded in
the dipole magnetic field of the central object,
can not have a Keplerian behavior far from the
plane $z = 0$.

The radial equilibrium of the disk configuration
corresponds to the following force balance
\begin{equation}
G\Mstar\rho \left[
-\frac{r\psi^3}{\muo^3} +
\frac{\psiD}{\muo r}\right] =
-\partial_rp - \frac{\partial_r\psi}{4\pi r}
\mD{\psi}
\, ,
\label{radeq}
\end{equation}
where $\rho=\rho(r,z^2)$ and $p=p(r,z^2)$ denote
the mass density and the thermostatic pressure,
respectively.

The vertical configuration equation reads as
\begin{equation}
-\partial_z p - \rho \frac{G\Mstar z}{\muo r^2}\psiD  
- \frac{\partial_z\psi}{4\pi r}\mD{\psi}
 = 0
\, ,
\label{vereq}
\end{equation}
where 
\begin{equation}
\mD{\psi}\equiv\partial_r\left(
\frac{1}{r}\partial _r\psi\right)
+\frac{1}{r}\partial _z^2\psi\,.
\end{equation}

\section{Perturbation Scheme}\label{sec:persc}

We now split the flux surface function $\psi$ as follows:
\begin{equation}
\psi = \psiD + \zeta,\hspace{5mm}
\mid \zeta \mid \ll \mid \psiD\mid\,.
\end{equation}
Here $\zeta(r,z^2)$ describes the
electromagnetic backreaction of the confined plasma.
Via the approximation
$\psi^3\simeq \psiD^3 + 3\psiD^2\zeta$, and 
recalling that the
toroidal currents, associated with $\psiD$
identically vanish,
the radial and vertical configuration equations rewrite as
\begin{subequations}\label{confsys1}
\begin{equation}
\begin{split}
\rho \frac{G\Mstar}{\muo} \left[-\frac{r}{\muo^{2}}\left(\psiD^3
+ 3\psiD^2\zeta\right) + 
\frac{\psiD}{r}\right] \\
=-\partial _rp - \frac{1}{4\pi r}
\left(\partial _r\psiD + \partial _r\zeta \right)\mD{\zeta}\,,
\label{radeq1}
\end{split}
\end{equation}
\begin{equation}
\partial _zp
+\rho
\frac{G\Mstar z}{\muo r^2}\psiD  
+ \frac{1}{4\pi r}
\left(\partial _z\psiD + \partial _z\zeta\right)\mD{\zeta}= 0 
\, ,
\label{vereq1}
\end{equation}
\end{subequations}
respectively.
In these equations, we retained the gradients
of $\zeta$, because, in agreement with the
\emph{drift ordering}, they can be relevant
and even dominant, despite the smallness of $\zeta$.

Taking into account the thin nature of the disk,
i.e., that its half-depth $H(r)$ satisfies the relation
$H(r)\ll r$, we can expand the vertical dependence
of the above equations in agreement with the approximation
$(r^2 + z^2)^a \simeq r^{2a}(1 + a z^2/r^2)$.
Hence, the configurational system \eref{confsys1} is recast as 
\begin{subequations}\label{confsys2}
\begin{equation}
\begin{split}
\frac{3 \rho  G\Mstar}{r^4}\left(z^2         
- \frac{(r^3 - 3rz^2)\zeta}{\muo}\right) =\\
=-\partial_r p - \frac{1}{4\pi r}
\left[-\frac{\mu _0}{r^2}\left(1 - \frac{9z^2}{2r^2}\right) 
+ \partial_r \zeta\right]\mD{\zeta}\,,
\label{radeq2}
\end{split}
\end{equation}
\begin{equation}
\partial _z p
+\rho
\frac{G\Mstar z}{r^3} 
+ \frac{1}{4\pi r}
\left(-\frac{3\mu _0z}{r^3} + \partial _z\zeta\right)
\mD{\zeta} = 0
\, .
\end{equation}
\end{subequations}
We now split the mass density $\rho$ and the 
pressure $p$ into the background (overbar) 
and perturbation (careted) components, as follows
\begin{equation}
\rho = \bar{\rho} + \hat{\rho}
\, ,\quad\quad \,
p = \bar{p} + \hat{p}
\, ,
\label{split}
\end{equation}
requiring that the quantities $\bar{\rho}$ and
$\bar{p}$ are linked via the isothermal
relation $\bar{p} = \vs^2\bar{\rho}$. We also determine
the form of $\bar{\rho} (r,z^2)$ by imposing the validity
of the gravothermal vertical equilibrium
\begin{equation}
\partial _z\bar{p} +\bar{\rho}\omegaKo^2 z = 0
\, \Rightarrow \,
\bar{\rho} = \rho_0(r)
\exp \left(-\frac{\omegaKo^{2} z^2}{2\vs^2}\right)
\, ,
\label{vereq11z}
\end{equation}
where  we remind that $\rho_0(r)$
denotes the mass density on the equatorial plane.
Observing that
\begin{equation}
\partial _r \bar{p} = \left(\frac{3G\Mstar z^2}{2r^4}
+ \frac{\vs^2}{\rho_{0}}\frac{d\rho_0}{dr}\right)
\bar{\rho}\, ,
\label{parp}
\end{equation} 
in order to restate the
radial and the vertical equilibria in a simpler form,
we are led to require that the following conditions hold 
\begin{equation}\label{condsd}
\!\!\!\frac{\zeta r}{\muo} =
\frac{\zeta}{\psiDo}\gg
\frac{3 z^2}{2 r^2}\,, \quad
\frac{z^2}{r^2} \ll 1
\, , \quad
\frac{\zeta}{\psiDo}\gg\gamma
\frac{r \vs^2}{3G\Mstar}
\, ,
\end{equation}
($\psiDo = \psiD(r, z=0)$).
Here we make the ansatz $\rho_{0}\propto r^{\gamma}, \gamma = \const$ (see below) and obtain the following system of equations from \erefs{confsys2}
\begin{subequations}\label{confsys3}
\begin{equation}
\label{radeq3}
\begin{split}
\left(\bar{\rho} + \hat{\rho}\right)
\frac{3G\Mstar}{\muo r} \left(1 - 3\frac{z^2}{r^2}\right)\zeta \\
=\partial _r\hat{p} +\frac{1}{4\pi r}
\left[-\frac{\muo}{r^2}\left(1 - \frac{9z^2}{2r^2}\right) 
+ \partial _r\zeta \right]
\mD{\zeta}\,,
\end{split}
\end{equation}
\begin{equation}
\partial_z \hat{p} 
+ \hat{\rho} \frac{G\Mstar z}{r^3} 
+ \frac{1}{4\pi r}
\left(-\frac{3\mu _0z}{r^3} + \partial _z\zeta\right)
\mD{\zeta} = 0
\, .
\label{vereq3}
\end{equation}
\end{subequations}

\subsection{Toward the crystalline structure}\label{sec:crstr}

In order to find the region of applicability
for the crystalline structure of the disk, outlined
in the local model of Coppi \cite{C05}, we impose some
specific restrictions.
Denoting by $k$ the wavenumber of the radial dependence
characterizing $\phi=\zeta/ \sqrt{r}$, we make the request
to live in the disk zone where $kr\gg 1$. Thus, 
under such a restriction and neglecting
the quantity $\partial_z \psiD$ in the vertical
force balance, the disk configuration is
determined via the following system
\begin{subequations}\label{toteq} 
\begin{equation}
\begin{split} 
\displaystyle\frac{3G\Mstar\rho_0(r)}{\muo\sqrt{r}}\left(\exp\left\{-\displaystyle\frac{\omegaKo^2 z^2}{2\vs^2}\right\}
+ \hat{D}\right)
 \left(1 - 3\frac{z^2}{r^2}\right)\phi 
\\
\\=\partial_r \hat{p} +\frac{1}{4\pi r}
\left[\partial_r \phi -\frac{\muo}{r^{5/2}}\left(1 - \frac{9z^2}{2r^2}\right)\right]
\left(\partial ^2_r\phi + \partial _z^2\phi  \right)\,,
\end{split} 
\end{equation}
\begin{equation}
\partial_ z\hat{p} + \hat{\rho} \displaystyle\frac{G\Mstar z}{r^3} 
+ \frac{1}{4\pi r}
\partial_z \phi
\left(\partial ^2_r\phi + \partial _z^2\phi\right) = 0\, , 
\end{equation}
\end{subequations}
where we defined
$\hat{D} = \hat{D}(r,z^2) \equiv \hat{\rho}/\rho_0$.
If we indicate by $h$ the characteristic scale
for the $z$-dependence of $\zeta$, the
possibility to neglect the term $\partial_z\psiD$
in the vertical equilibrium, relies on the
validity of the requirement
$\zeta /\psiD\gg 3 zh/r^2$.

\section{The linear case}\label{sec:lincas}

\begin{figure}[ht]
   \centering
   \includegraphics[width=\columnwidth]{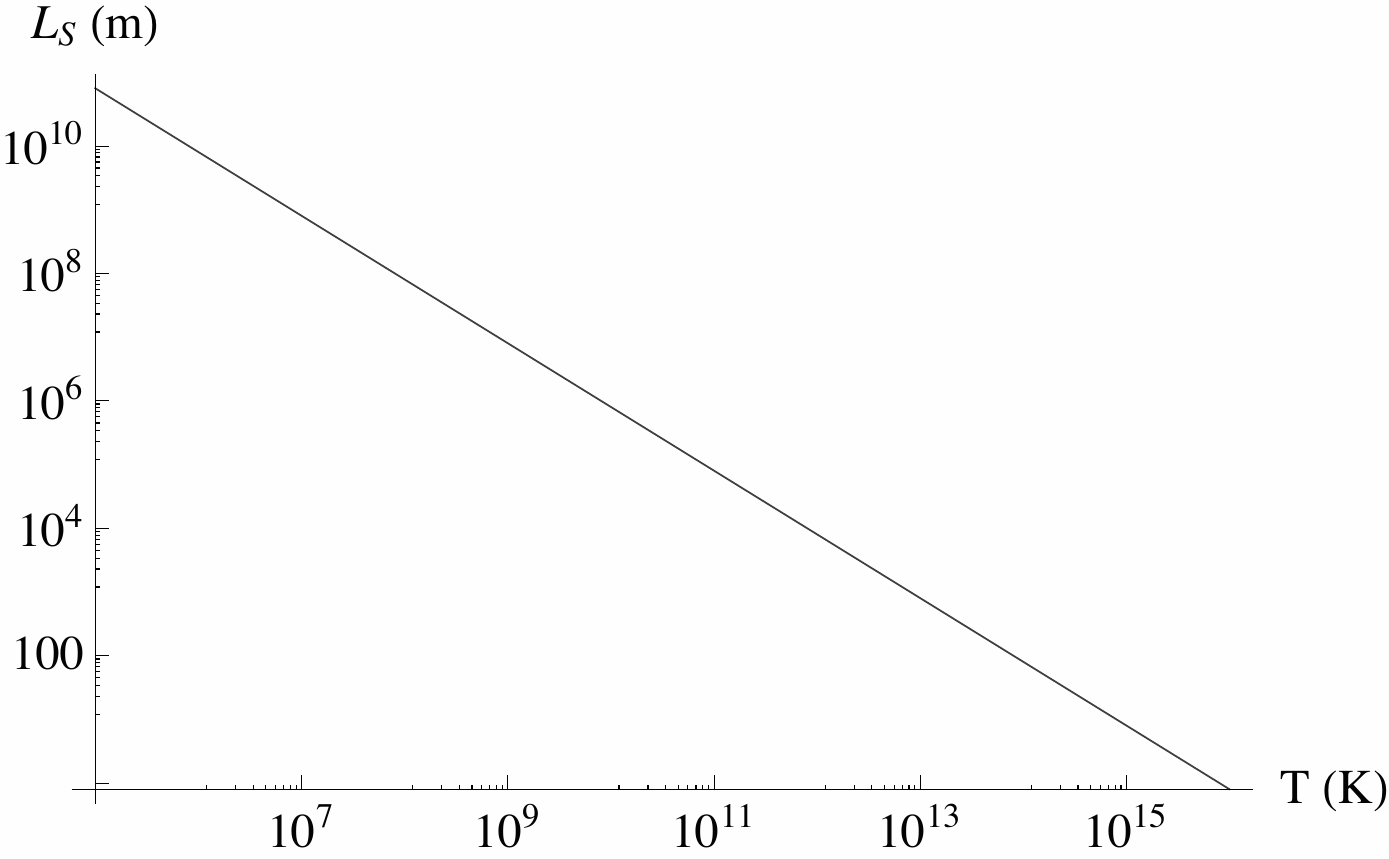} 
   \caption{The parameter $\Ls$ as a function of the temperature $T$ of the isothermal disk for a $\Mstar=2M_\odot$}
   \label{fig:Ls}
\end{figure}

\begin{figure}[ht]
   \centering
     \includegraphics[width=\columnwidth]{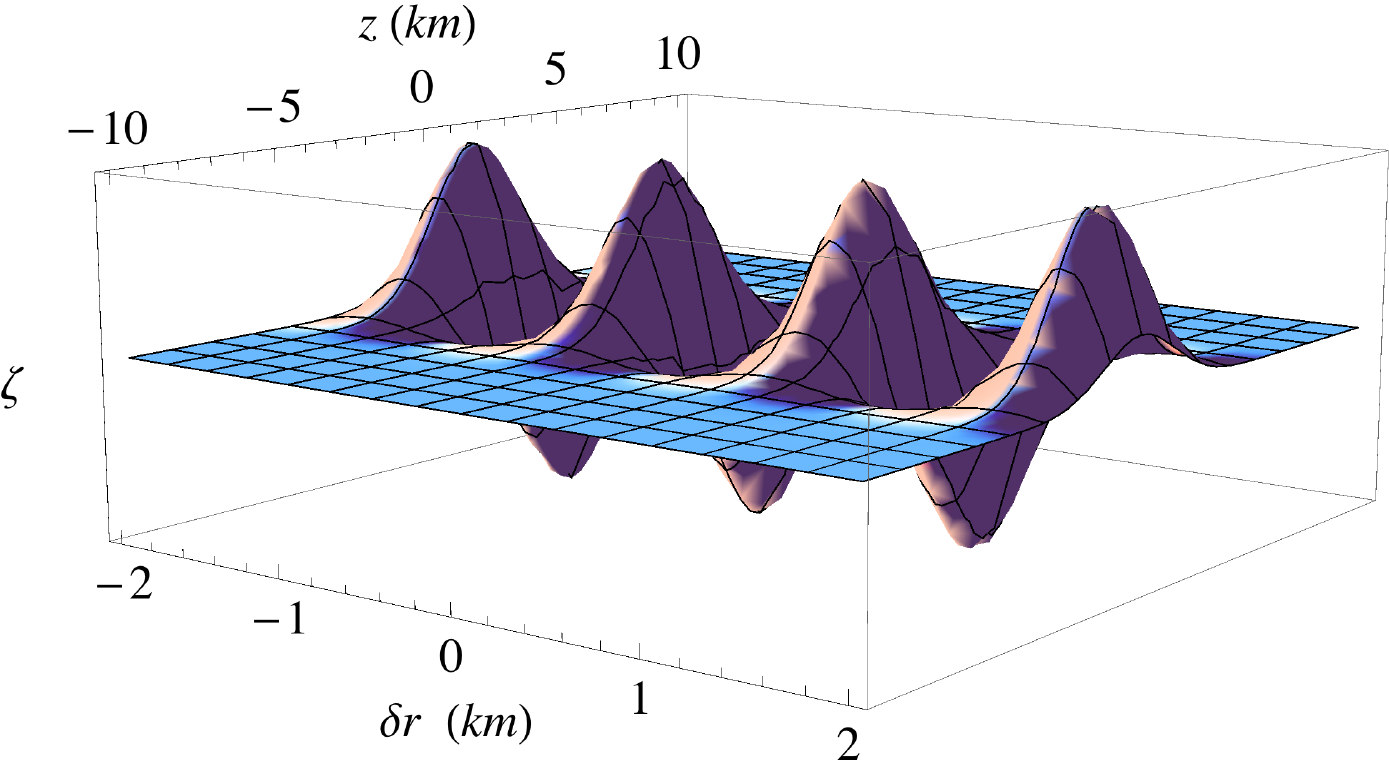} 
   \caption{Local oscillating behavior of the perturbed flux function $\zeta$ around a radius $r=10^3 \km$. The chosen parameters are: $A=1$, $T=10^7$K, $\Ls\simeq8\times10^5$Km, $B=10^{12}$Gauss and $m=0.02M_\odot$}
   \label{fig:zeta3D}
\end{figure}

\begin{figure}[ht]
   \centering
   \includegraphics[width=\columnwidth]{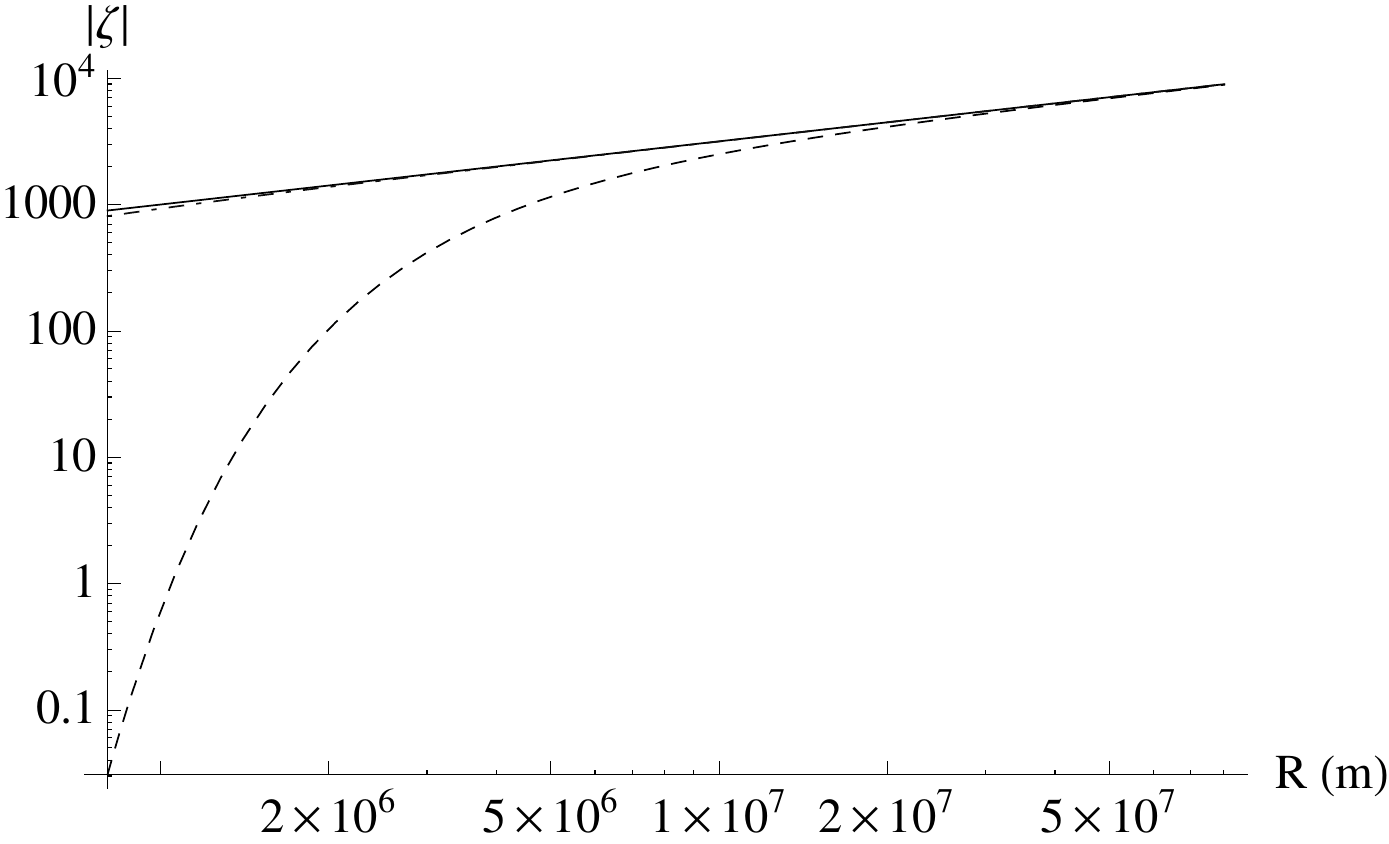} 
   \caption{Global profile of the perturbed flux function $\zeta$ for fixed values of $z=0, 10^2$ and $10^3$m (clearly, the oscillating factor $\sin(kr)$ has been omitted). The chosen parameters are the same as in \fref{fig:zeta3D}. The weak radial dependence of the amplitude is evident.}
   \label{fig:zeta}
\end{figure}

We now study the linear case, corresponding
to a sufficiently low electromagnetic backreaction
in the plasma, so that we can neglect non-linear
terms in $\phi$ when fixing the equilibrium.
For instance,
$\partial _r\psiD\gg \partial _r\zeta$ is
equivalent to the condition
\begin{equation}\label{condlin}
\frac{\zeta}{\psiD}\ll \frac{1}{kr}\, .
\end{equation}
Furthermore, we observe that the background
mass density can be expanded as
\begin{equation}\label{exprho}
\bar{\rho} \simeq \rho_0(r) \left(1 - \frac{G\Mstar z^2}{2r^3\vs^2} \right) =
\rho_0(r) \left[1 - \left(\frac{\rs c^2}{4r\vs^2}\right)\frac{z^2}{r^2}\right]\, ,
\end{equation}
where $\rs \equiv 2G\Mstar/c^2$ is the
Schrwarzshild radius of the central object.
In what follows, we assume to be in that region
of the disk where the condition
\begin{equation}\label{pippo}
\displaystyle\frac{\rs}{4r} \gg \displaystyle\frac{\vs^2}{c^2} \,,
\end{equation}
holds and we
can approximate,
in this linear regime, the configurational
system \reff{toteq} as
\begin{subequations}\label{confsys4}
\begin{equation} 
\label{toteq1} 
\rho_0
\displaystyle\frac{3G\Mstar}{\muo} \left(1 - \displaystyle\frac{\Ls z^2}{r^3}\right)\phi +
\frac{\muo}{4\pi r^3}\left(\partial^2_r\phi + \partial_z^2\phi\right) = 0\,,
\end{equation}
\begin{equation}
\partial _z\hat{p}
+\hat{\rho} \displaystyle\frac{G\Mstar z}{r^3} = 0\,,\label{toteq2}
\end{equation}
\end{subequations}
where 
\begin{equation}\label{defLs}
\Ls \equiv G\Mstar/2\vs^2\,, 
\end{equation}
(see \fref{fig:Ls}).
In these equations we neglected $\hat{D}$,
because $\bar{\rho}\gg \hat{\rho}$  in the linear regime, and also the
radial pressure gradient, in comparison to the
Lorentz force. 
We observe that condition \reff{pippo} reads as $r\ll \Ls$ and that it is implied by condition \reff{condsd}.
\Eref{toteq2} tells us that the perturbations follow the background
behavior, while the corresponding radial equilibrium \reff{toteq1} has an intriguing feature: if we take
$\rho_0 = m/r^3$, with $m = \const.$ and
define $k^2 = 12\pi G\Mstar m/\mu _0^2$, then
we arrive at the radial equilibrium equation in the form
\begin{equation}
\left(\partial^2_r\phi + \partial_z^2\phi\right) = 
-\left(1 - \frac{\Ls z^2}{r^3}\right)
k^2\phi 
\, .
\label{radtoteq1} 
\end{equation}
In the considered limits $z/r\ll 1$, $kr\gg 1$
and for $r\gg (\Ls/k^2)^{1/3}$
(and remembering $r\ll \Ls$), the
solution of the equation above takes the form
\begin{equation}
\phi(r,z^2) = A\sin (kr)\exp \left\{-\displaystyle\frac{k\sqrt{\Ls}z^2}{2r^{3/2}} \right\}\, ,
\label{crysf}
\end{equation}
$A$ being a constant amplitude.
This oscillating form of $\phi$, i.e.,
of $\zeta = \sqrt{r}\phi$ restores, under the
set of conditions we fixed, the crystalline
structure emerging in the local model of Coppi. The local oscillating behavior of $\zeta$ is sketched in \fref{fig:zeta3D}, while the global behavior is described in  \fref{fig:zeta}.

All the conditions imposed on the global model to recover the linearized radial oscillating behavior of the magnetic surface function, can be summarized in the following three physical restrictions on the considered disk model: 
\begin{enumerate}
\item The disk is assumed to be thin, i.e., $z^2/r^2\ll 1$.
\item The wavelength of the perturbations, due to the electromagnetic backreaction, is much smaller than the disk size, i.e., $kr\gg 1$.
\item The disk configuration must be restricted to the radial region where the condition $r\ll \Ls$ holds, i.e.,
\begin{equation} 
r\ll \frac{G\Mstar}{2\vs^2} = \frac{\rs c^2}{4\vs^2} = \frac{\mi c^2}{8 \Kb T^{2}}\rs \, , \label{funineq} \end{equation}
where, the sound velocity $\vs$ has ben expressed via the disk temperature $T$.
\end{enumerate}

In this scheme, the existence of the crystalline structure requires that the disk temperature not exceed a given value. For instance, for a neutron star of about two Solar masses and a disk size of $10^6 \km$, in order to apply our model over the whole disk, we would get the condition $ \Kb T\ll 1 \keV$. Such an estimation leads us to infer that the present result, as that obtained in \cite{C05}, remains valid for a class of rather cold disks.

\section{Phenomenological Implications}\label{sec:phenomenological}

Let us analyze the physical hints that we can get from
the analysis of the global model for the crystalline
structure, as traced above.

In order to compare the radial scaling of the perturbations
to the size of the allowed region for the crystalline
structure, we have to consider the ratio between the wavelength
$\lambda = 2\pi /k$ and the typical scale $\Ls$, i.e.,
 \begin{equation}
 \frac{\lambda}{\Ls} = \sqrt{\frac{\pi \muo^2}{3G\Mstar m}}
 \frac{2\vs^2}{G\Mstar} = \displaystyle\sqrt{\frac{32\pi}{3}}
\displaystyle\frac{\vs^2}{c^2}\sqrt\frac{\ell^3}{\rs^3}
\, ,
\label{ratiofu}
 \end{equation} 
where we defined the characteristic length
 \begin{equation}\label{ratiolu}
\ell \equiv \sqrt[3]{\frac{\muo^2}{mc^2}}\, .
\end{equation}
Since $\vs <  c$ and $\ell \ll \rs$
(for a typical neutron star we have value in the
range $\sim 10$m; see below for more details), the above ratio is many orders of
magnitude smaller than unity.  The
oscillation scale of the perturbations, therefore, lives in the
microscales of the system, very much below the
characteristic length $\Ls$.
The most relevant phenomenological implication of
this issue, is in the impact that
the crystalline structure can have on the global configuration
of the disk. The very small characteristic length of the magnetic flux
surface oscillations prevents the ring sequence
emerging in the non-linear regime (see \sref{sec:extreme}) from being on a macroscopic
scale, and so able to account for certain discontinuous
structures resulting from the observations of real
astrophysical compact objects. Such kind of macroscopic structures can take
place in this framework in limiting situations only, i.e., when the parameter
$m$ is particularly small and $\muo$ is sufficiently large. However, this consideration
does not affect the relevance of the existence of such
microstructures toward the global equilibrium profile.
In fact, this microscopic nature of the crystalline structure
and of the ring sequence,
plays a crucial role in fixing the fundamental instabilities
characterizing the disk profile. In particular,
the small-scale radial oscillations can trigger instabilities
having very different morphologies with respect to the
MRI, as already discussed for the local configuration
in \cite{Coppi:2008p98}. Thus, more than through 
direct observations in the optical or X-ray bands,
we expect that the existence of such structures inside
the disk can be revealed by specific features of the  triggered turbulent
regimes that can be unveiled in the resulting emission processes.
The role that the microscopic crystalline structure
can play in the onset and in the establishment of a
turbulent scenario, can deeply influence the
transport processes inside the disk plasma,
especially in view of energy-momentum transfer from the
micro- to the meso-scales of the system.

Another relevant phenomenological question concerns
the restricted region where the crystalline
profile and the fragmentation of the disk in a ring series
can take place. Indeed, the condition
$r\ll \Ls$ seems to confine the radial domain
allowed for the magnetic flux surface oscillations
in an internal portion of the thin disk.
From a phenomenological point
of view, this inequality is sensitive to
the accretion system parameters, i.e., the mass of the central
object and the temperature of the disk, so identifying which fraction of the plasma extension
\footnote{Such fraction is roughly proportional to the ratio $\Ls/R_\mathrm{ext}$, where $R_\mathrm{ext}$ denotes the external
radius of the disk configuration.
}
is involved in the crystalline profile.
In particular, for configurations extended enough,
having very large values of $R_\mathrm{ext}$ and 
whose plasma is significantly hot
(so that the sound velocity is a few orders of magnitude
smaller than the speed of light), the
restriction above can limit the small scale radial
oscillations to a very tiny portion of the accreting
plasma. However, it is possible to characterize the class
of disks to which such restrictions can be applied
in a more simple and meaningful way, which is also
consistent with the approximation scheme adopted
above.
In fact, the restriction $r\ll \Ls$
can be easily restated as follows
\begin{equation}
r\ll \Ls = \frac{G\Mstar}{2\vs^2} =
\frac{\omegaK^2(r)}{2\vs^2}r^3 =
\frac{r^3}{H^2(r)}
\, ,
\label{condri}
\end{equation}
where we recall that $H(r)$ is the
effective half-depth of the isothermal disk
(see \eref{vereq11z}).
Thus, the restriction
$r\ll \Ls$ is equivalent to the requirement that
the disk has a real thin profile, i.e.,
$H(r)\ll r$. Realizing such an equivalence
allows us to join together two of the fundamental
constraints at the ground of our derivation of the
global model, i.e., $r\ll \Ls$ and $z/r \ll 1$.
In fact the former, being equivalent to $H(r)\ll r$
automatically ensures the validity of the latter.
It is worth noting, however, that for a thick disk,
while the constraint $r\ll \Ls$ is necessarily violated
(with non-trivial implications for the existence of
a global crystalline profile), the condition
$z\ll r$ can still be satisfied in the proximity
of the equatorial plane of the axisymmetric configuration.

We conclude this section by stressing how the
two fundamental requirements we relied on when deriving the
radial oscillation of the disk morphology, overlap
the fundamental hypotheses of dealing with a thin disk
in which the plasma backreaction is a small scale phenomenon,
namely, $H(r)\ll r$ and $kr\gg 1$, respectively.
This issue makes our analysis fully consistent
with the ideas introduced in \cite{C05,CR06} and
substantiates the guess that the crystalline morphology
and the ring sequence scenario are very general
features of an accreting plasma well
confined close to the equatorial plane.
However, our study also has the merit to outline
the microscopic nature of the backreaction scale
(more than a simple small scale behavior postulated
by Coppi), so opening a precise direction
in the understanding of the role that such a radial
periodicity of the plasma can play in the establishment of
significant processes of transport of matter and
angular momentum across the disk.

With respect to the estimation of the characteristic length
$\ell$, it is worth noting the following relation (valid in the linear case)
between the total mass of the disk $\Md$ and the
parameter $m$ entering the equatorial distribution
$\rho_0(r) = m/r^3$, i.e.,
\begin{equation}\label{mdm}
\begin{split}
\Md &= 2\pi m\int_{R_\mathrm{int}}^{R_\mathrm{ext}}\int_{-H(r)}^{H(r)}
 \frac{1}{r^2}e^{- {z^2}/{H^2(r)}} dz dr\\
& = 8\pi^{3/2} m\, \erf(1)\displaystyle \sqrt{\frac{\Kb T}{G\Mstar \mi}} \left(R_\mathrm{ext}^{1/2} - R_\mathrm{int}^{1/2}\displaystyle \right)\\
&\simeq 1.8 \times 10^3 m \displaystyle\sqrt{\frac{\Kb T}{\keV}} \, .
\end{split}
\end{equation}
Here $\erf(x)=2 \pi^{-1/2} \int_0^x e^{-t^2} \, dt$ is the error function, $\Mstar=2\Msun$ is the mass of the central object, and the last equality stands for an inner and external disk radii of $R_\mathrm{int} = 10^3\km$ and $R_\mathrm{ext}=10^6 \km$ respectively.
As a result we find that the parameter $m$ explicitly depends on the temperature $T$ and on the mass of the disk $\Md$ via the relation
\begin{equation}\label{mmd}
m= \frac{5.5 \times 10^{-5} \Md}{\sqrt{\frac{\Kb T}{\keV}}}\,.
\end{equation}
Assuming that $\Md$ takes values in the range $[10^{-5},10^{-2}] \Msun$, $\Kb T$ may vary within $[10^{-1},10^2] \keV$, the stellar radius to be equal to  $10 \km$, the parameter $m$ assumes values in the interval $[5.5\times 10^{-11},1.7\times 10^{-5}]\Msun$.

\begin{figure}[ht]
   \centering
   \includegraphics[width=\columnwidth]{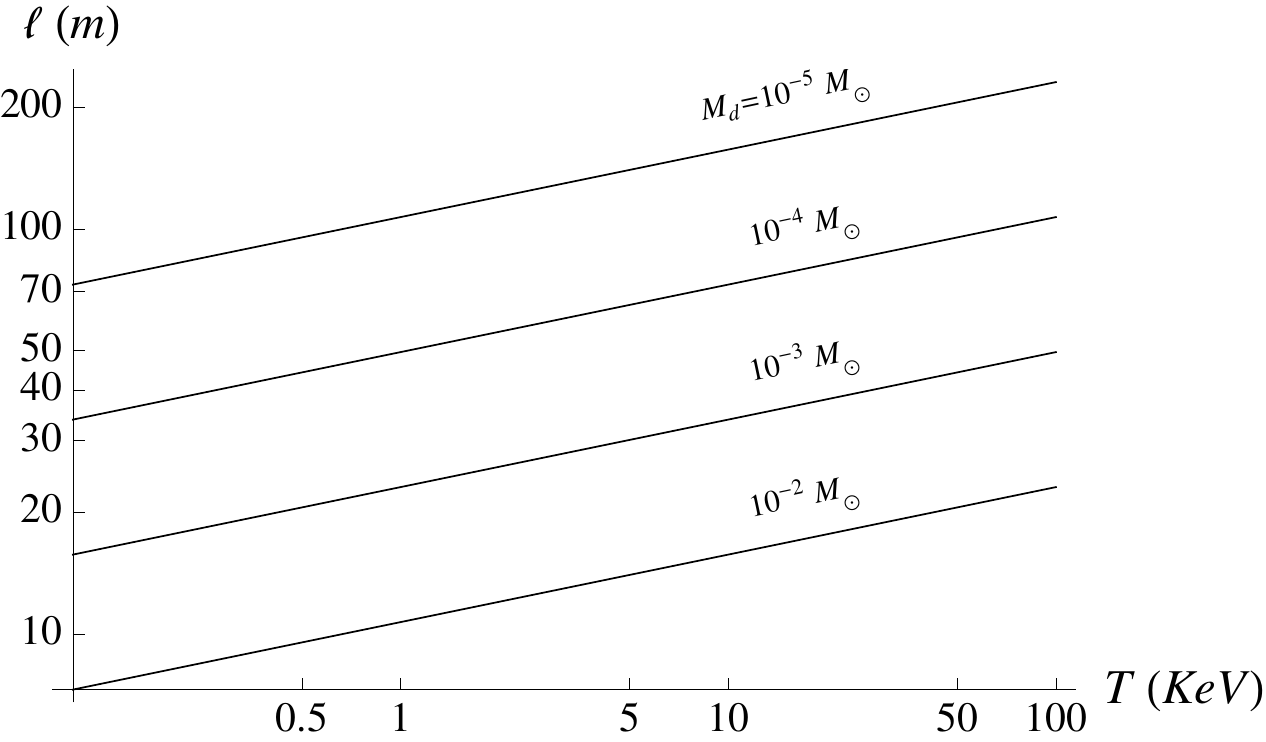} 
   \caption{Behavior of the characteristic length $\ell$ as a function of the temperature $T$ of the disk. The different values above each single plot denotes the mass of the accretion disk $\Md$ in Solar mass units for which \eref{ratiolunum} is evaluated. It is worth noting how such a characteristic length is always much smaller than the typical dimensions of a disk.}
   \label{fig:ell}
\end{figure}

\begin{figure}[ht]
   \centering
   \includegraphics[width=\columnwidth]{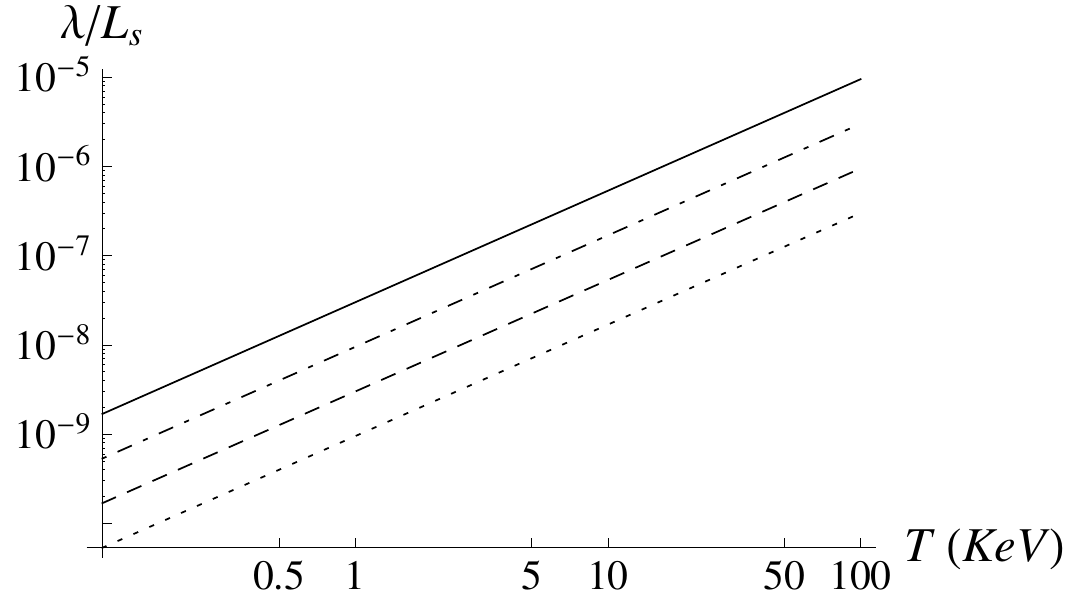} 
   \caption{Plot of the $\lambda/\Ls$ ratio as a function of the temperature $T$ of the disk. The different curves correspond to different values of $\Md/\Msun$  (from top to bottom: $10^{-5}, 10^{-4}, 10^{-3}, 10^{-2}$) for which \eref{ratiofu} is evaluated.}
   \label{fig:lambdals}
\end{figure}

The parameter $\ell$ given by \eref{ratiolu}, then, is related to the parameters of the model as follows
\begin{equation}\label{ratiolunum}
\ell \simeq 2.3 \left(\displaystyle\frac{B}{\mathrm{10^9 G}}\right)^{\frac{2}{3}}
   \displaystyle\left(\displaystyle\frac{\Kb T}{\keV}\right)^{\frac{1}{6}}\left(\frac{\Md}{\Msun}\right)^{-\frac{1}{3}} \mathrm{cm}\,.
\end{equation}
For different values of the mass of the accretion disk $\Md$ and  for a $B$-field of $10^{12}$ Gauss, the behavior of $\ell$ and the ratio $\lambda/\Ls$ $\reff{ratiofu}$ as functions of the temperature are given in \fref{fig:ell} and \fref{fig:lambdals} respectively. From the behavior of $\ell$, it is evident that the condition $\ell\ll \rs$ is always verified.

\section{Extreme non-linear Regime}\label{sec:extreme}

We now analyze the opposite case with respect to the linear regime, when the electromagnetic backreaction dominates the background dipole contribution
(i.e., $\partial _r\zeta \gg \partial _r\psiD$)
and the perturbed mass density $\hat{\rho}$ is larger than $\bar{\rho}$ fixed by the gravostatic equilibrium \reff{exprho}; more precisely, we can apply the following hierarchy relations
\begin{equation}\label{condnl} 
\frac{\zeta}{\psiDo} \gg \frac{1}{kr}\,, \quad  \quad \hat{\rho} \gg \bar{\rho} \, .
\end{equation}
Therefore, we can also require that the gravitational term of the system (\ref{toteq}) be negligible with respect to the pressure gradient, i.e., (here we are assuming $\hat{p}= \vshq\hat{\rho}$)
\begin{equation} \label{vertcond}
\displaystyle\frac{r}{\Lsh} \gg \frac{2hz}{r^2} \, ,
\end{equation}
where $\Lsh\equiv G\Mstar/2\vshq$.
We now split the pressure term in analogy to the surface function $\zeta$, i.e., $\hat{p}(r,z^2) = \hat{q}(r,z^2)/r$, so that
\begin{equation} \label{presappro}
\partial _r\hat{p} = \displaystyle\frac{1}{r}\left( \partial _r\hat{q}
- \frac{1}{r}\hat{q}\right) \simeq \frac{1}{r}\partial _r\hat{q} \, ,  
\end{equation}
valid for $kr\gg 1$. Hence, fixing the mass density on the equatorial plane in the form $\rho _0 = m^{\prime}/\sqrt{r}$, the configuration system \reff{toteq}  reads as
\begin{subequations}\label{toteqnl} 
\begin{eqnarray} 
&\displaystyle\frac{3G\Mstar m^{\prime}}{\mu _0} \hat{D}\phi = \partial _r\hat{q} + \frac{1}{4\pi} \partial _r\phi \Delta \phi&\,,\label{toteqnl1}  \\ 
&\partial _z\hat{q} + \displaystyle\frac{1}{4\pi } \partial _z\phi \Delta \phi = 0 \, .& \label{toteqnl2}  
\end{eqnarray}
\end{subequations}
In order to analyze the system above, let us
introduce the following dimensionless quantities
\begin{eqnarray}
&x = \tk r\,, \quad  \quad u = \displaystyle\frac{z}{h}\,,\quad  \quad
\xi = \frac{1}{(\tk h)^2}\,,\nonumber\\
\displaystyle&S = \displaystyle\frac{\hat{q}}{\muo^2 \tk^5}\,,\quad\quad
\tk = \left( \frac{3 G\Mstar m}{\muo^2L^{5/2}}
\right)^{2/9}\,,\label{dimless}\\
&\Phi = -\displaystyle\frac{\phi }{\muo \tk^{3/2}}
\, ,\nonumber
\end{eqnarray}
where we set $m^{\prime} = \tm/L^{5/2}$,
with $L = \const$.
By such definitions, system \reff{toteqnl}
rewrites in the form
\begin{subequations}
\begin{equation}
\hat{D}\Phi +
\partial _xS
+ \displaystyle\frac{1}{4\pi }
\partial _x\Phi
\left(\partial^2_x\Phi + \xi \partial^2_u\Phi\right) = 0\,, \label{toteqnlxx1}
\end{equation}
\begin{equation}
\partial _uS
+ \displaystyle\frac{1}{4\pi}
\partial _u\Phi
\left(\partial^2_x\Phi + \xi \partial^2_u\Phi \right) = 0\, .
\label{toteqnlxx2}
\end{equation}
\end{subequations}
We now solve these two equations in close
analogy to what is done in \cite{CR06} in this
same extreme non-linear regime.
We consider a function $\Phi$ in the form
$\Phi(x,u^2) = N(x)F(u^2)$, where $N$ is
an odd function of the radial variable.
In the limit $\xi\rightarrow 0$, which we are
focusing on,
from \eref{toteqnlxx2}, we easily get
\begin{equation}\label{asdfd}
S(x,u^2) = - \displaystyle\frac{1}{8\pi}F^2N\frac{d^2N}{dx^2}\, .
\end{equation}
Substituting this expression into \eref{toteqnlxx1} and setting
$\hat{D}(x,u) = K(x)F(u^2)$, we eventually get
\begin{equation}
\label{asdfdqw}
8\pi K(x) = \displaystyle\frac{d^3N}{dx^3}
- \frac{1}{N}\frac{dN}{dx}\frac{d^2N}{dx^2}\, ,
\end{equation}
where we must necessarily require $K\ge 0$,
to ensure the positivity of the mass density $\hat{\rho}=\hat{D} \rho_0$.

\begin{figure}[ht]
   \centering
   \includegraphics[width=\columnwidth]{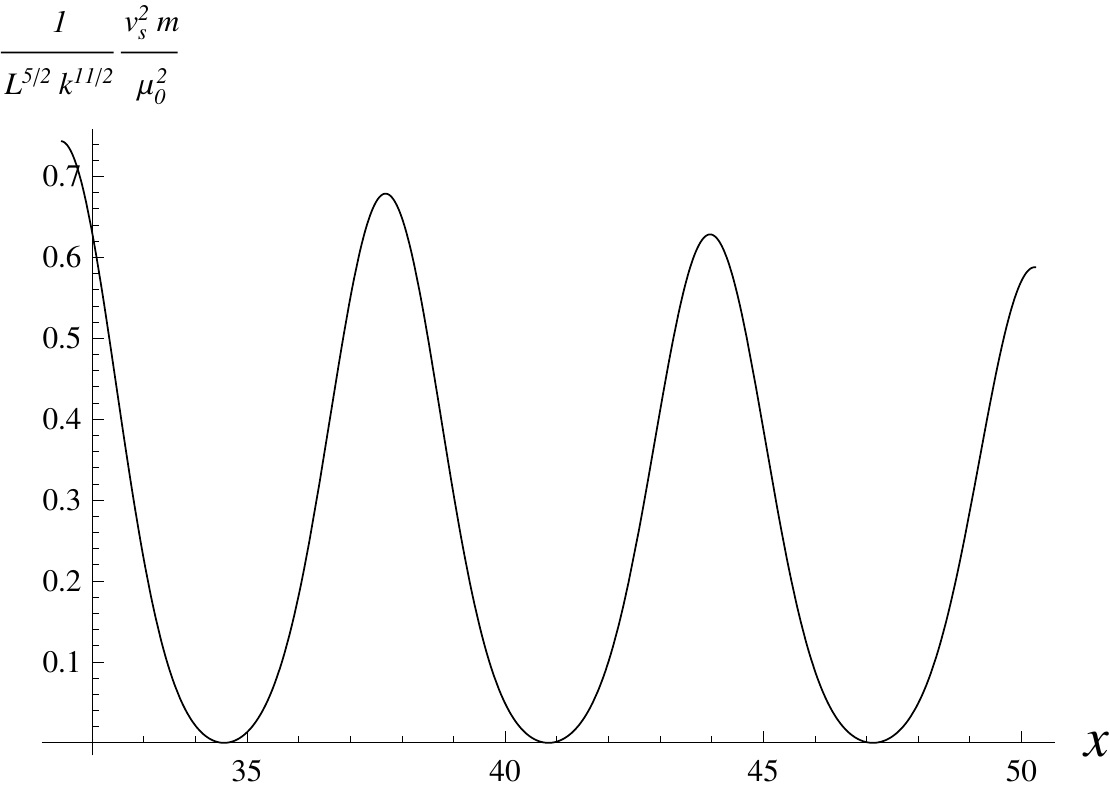} \\
   \includegraphics[width=\columnwidth]{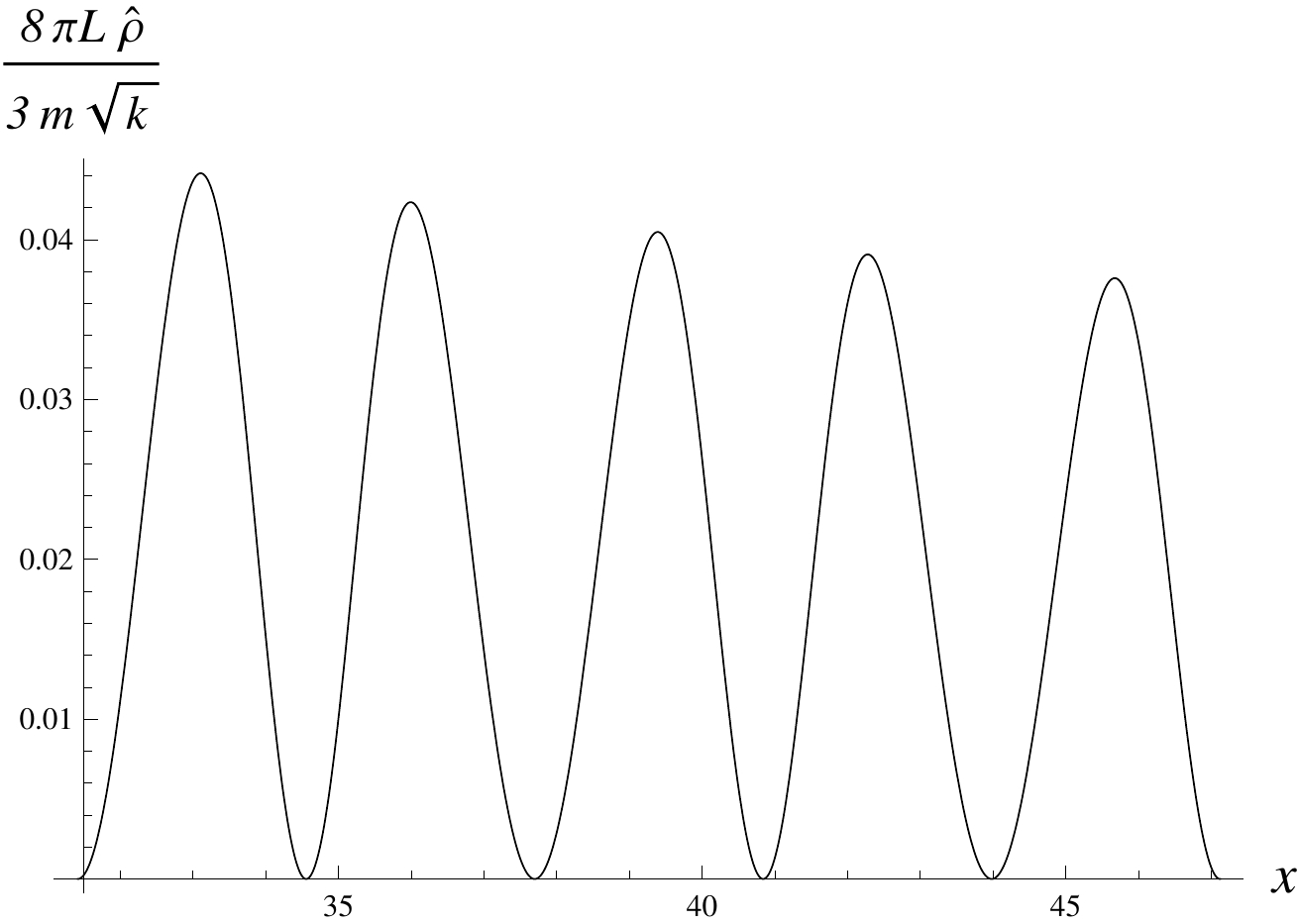} 
   \caption{Adimensional sound velocity $\vshq$ (top) and perturbed mass density $\hat{\rho}$ (bottom) for the case $A=1$ and $B=1/8$, and $F(u^2)=1$.}
   \label{fig:hatrho}
\end{figure}

If we assume the following simple form for $N(x)$
\begin{equation}
\label{ksin}
N(x) = A\left[\sin(x) + B \sin(2x)\right]\,,
\end{equation}
then
\begin{equation}
\label{ksincalcolata}
8\pi K(x) = \frac{6 A B \sin ^2(x)}{2 B \cos (x)+1}\,.
\end{equation}
Finally, we can calculate the quantity $\vshq$
via the relation
\begin{equation}
\begin{split}
\vshq &= \frac{\hat{p}}{\hat{\rho}} =
\displaystyle\frac{\muo^2 k^{11/2}  L^{5/2} }{m}\times\\
\times& \frac{A F\left(u^2\right) \left[2 B \cos (x)+1\right]^2 \left[8 B \cos (x)+1\right]}{6 B \sqrt{x}}\, .
\label{sjau}
\end{split}
\end{equation}
Since we have to require $K(x)$ and $\vshq$ to be positive definite, then parameters $A$ and $B$ must satisfy the following conditions
\begin{equation}
\label{condizioniAB} 
A>0\,,\hspace{5mm}0<B\leq\displaystyle\frac{1}{8}\,.
\end{equation}
For the particular choice $A=1$ and $B=1/8$, and $F(u^2)=1$, the (effective) perturbed  sound velocity $\vshq$ \reff{sjau}
and the perturbed mass density 
\begin{equation}
\hat{\rho }=\hat{D} \rho _0=\frac{K(x) \tm}{L^{5/2} \sqrt{r}}=\frac{3 \tm
   \sin ^2(x)}{8 \pi  L^{5/2} \sqrt{x/k} \left[\cos (x)+4\right]}\,,
 \end{equation}
are depicted in \fref{fig:hatrho}. The radial dependence of $\vshq$ can be interpreted as a corresponding behavior for the temperature perturbation induced by the plasma back-reaction. For sufficiently small scales of the crystalline structure, this temperature contribution becomes the dominant one and, in this scheme, the disk acquires a non-isothermal profile. 

We stress how, even in this global model, the emergence of a ring-sequence decomposition of the disk takes place as far as the induced magnetic field dominates the central object one, i.e., when the perturbations have a sufficiently small scale such that $\partial_r\zeta\gg\partial_r\psiD$. This fact confirms that a thin disk admits a plasma configuration characterized by microstructures which have the form of double and opposite current filaments. The instability properties of such radial profile of the disk is expected to deeply influence the averaged transport features of the global structure, especially in view of energy and angular momentum transfer from the micro- to the mesoscales of the system.

\section{Concluding Remarks}\label{sec:conrem}

We considered a two-dimensional axisymmetric ideal MHD 
model to describe the structure of a plasma disk
surrounding a compact astrophysical object. We neglected the self-gravity of the disk and
treated the central gravitational field as a
Newtonian profile induced by the mass of the
compact body. The magnetic field external to
the plasma is described by an exact dipole-like
structure associated with the intrinsic features
of the central astrophysical source.
We then settled the equilibrium configuration
of the plasma by accounting for the internal
electromagnetic back-reaction, described
via static perturbations of the magnetic
flux surfaces and of the thermodynamical quantities.
This way, we fixed the
global model associated with the analysis pursued in \cite{C05,CR06}, where a crystalline profile and a ring sequence for the disk emerged as the result of local configurations confined around fixed values of the radial coordinate.
In this respect, we neglected the poloidal component of the matter
flux, retaining the disk rotation as the only
dominant effect. Then the drift ordering was taken into account when splitting the configuration
variables into background and perturbation components.
The possibility to link the present global model
to the local configuration is offered by
a suitable radial scaling of the amplitudes
of the oscillating-like behaviors emerging
at a fixed radius; furthermore our scheme 
relies on the small scale nature of the perturbation in comparison to the disk size. We were able both in the linear and in the
extreme non-linear cases, discussed above in detail,
to recover the corresponding local equilibrium
profiles. The amplitude scaling of the
perturbed magnetic flux surfaces, mass density
and pressure were properly fixed. These results show
that the crystal structure of the magnetic field
and the ring-like profile of the mass density are rather general features of the
two-dimensional, axisymmetric, ideal MHD model.

The derivation of the radial scaling
requires, however,  a certain number of restrictions
on the fundamental quantities involved in
setting the plasma distribution. A key
request is that the plasma (suffering this
decomposition in periodic substructures) must be
confined within a certain region from the
central object, much smaller than the characteristic
length $\Ls$ defined in \eref{defLs}.
In the case of the most common stellar sources,
this request suggests that the crystal morphology
of the disk be preferably expected in relative
low-temperature plasma disks, while the case of a
very hot disk profile seems to require a very
large value of the central mass. For instance, in
the case of plasma structures surrounding the very
massive black holes at the center of Active Galactic
Nuclei, the central mass is typically
of the order $10^8-10^9$ Solar masses, leading to very high
values of the length $\Ls$. However both in the stellar and in the
galactic context, the internal nature of the
crystal profile, located sufficiently close
to the central object, suggests the necessity
for General Relativistic effects in describing the
gravitational interaction.
For a discussion of the role of the present
configuration of plasma and its instabilities
in the Active Galactic Nuclei morphology and for the
necessity to include relativistic corrections,
see \cite{CoppiHighEnergy}. Finally we stress how in \sref{sec:phenomenological}, we outlined that such restrictions on the radial region where the crystalline structure may take place is at all equivalent to requesting the thinness of the disk. In this sense, it is immediate to recognize the class of accretion disks to which the obtained result can apply, i.e., all those configurations for which the sound velocity is much smaller than the rotational flow of the disk.

This work was developed within the framework of the CGW Collaboration (http:// www.cgwcollaboration.it). R.B. acknowledges ICRA (International Center for Relativistic Astrophysics) for having partially supported this work.

\newcommand{\grg}{Gen. Rel. Grav.}

\newcommand{\araa}{Annual Review of Astronomy \& Astrophysics}
\newcommand{\mnras}{Mon. Not. R. Astron. Soc}
\newcommand{\aap}{A.\&A.}
\newcommand{\epl}{Europhys. Lett.}
\newcommand{\ppcf}{Plasm. Phys. Contr. Fus.}
\newcommand{\PP}{Phys. Plasmas}
\newcommand{\SA}{Soviet Astron.}
\newcommand{\nar}{New Astron. Rev.}
\newcommand{\physrep}{Phys. Rep.}

\bibliography{DischiFinale}

\begin{thebibliography}{27}%
\makeatletter
\providecommand \@ifxundefined [1]{%
 \@ifx{#1\undefined}
}%
\providecommand \@ifnum [1]{%
 \ifnum #1\expandafter \@firstoftwo
 \else \expandafter \@secondoftwo
 \fi
}%
\providecommand \@ifx [1]{%
 \ifx #1\expandafter \@firstoftwo
 \else \expandafter \@secondoftwo
 \fi
}%
\providecommand \natexlab [1]{#1}%
\providecommand \enquote  [1]{``#1''}%
\providecommand \bibnamefont  [1]{#1}%
\providecommand \bibfnamefont [1]{#1}%
\providecommand \citenamefont [1]{#1}%
\providecommand \href@noop [0]{\@secondoftwo}%
\providecommand \href [0]{\begingroup \@sanitize@url \@href}%
\providecommand \@href[1]{\@@startlink{#1}\@@href}%
\providecommand \@@href[1]{\endgroup#1\@@endlink}%
\providecommand \@sanitize@url [0]{\catcode `\\12\catcode `\$12\catcode
  `\&12\catcode `\#12\catcode `\^12\catcode `\_12\catcode `\%12\relax}%
\providecommand \@@startlink[1]{}%
\providecommand \@@endlink[0]{}%
\providecommand \url  [0]{\begingroup\@sanitize@url \@url }%
\providecommand \@url [1]{\endgroup\@href {#1}{\urlprefix }}%
\providecommand \urlprefix  [0]{URL }%
\providecommand \Eprint [0]{\href }%
\providecommand \doibase [0]{http://dx.doi.org/}%
\providecommand \selectlanguage [0]{\@gobble}%
\providecommand \bibinfo  [0]{\@secondoftwo}%
\providecommand \bibfield  [0]{\@secondoftwo}%
\providecommand \translation [1]{[#1]}%
\providecommand \BibitemOpen [0]{}%
\providecommand \bibitemStop [0]{}%
\providecommand \bibitemNoStop [0]{.\EOS\space}%
\providecommand \EOS [0]{\spacefactor3000\relax}%
\providecommand \BibitemShut  [1]{\csname bibitem#1\endcsname}%
\let\auto@bib@innerbib\@empty
%</preamble>
\bibitem [{\citenamefont {{Coppi}}(2005)}]{C05}%
  \BibitemOpen
  \bibfield  {author} {\bibinfo {author} {\bibfnamefont {B.}~\bibnamefont
  {{Coppi}}},\ }\href@noop {} {\bibfield  {journal} {\bibinfo  {journal} {\PP}\
  }\textbf {\bibinfo {volume} {12}},\ \bibinfo {pages} {7302} (\bibinfo {year}
  {2005})}\BibitemShut {NoStop}%
\bibitem [{\citenamefont {{Coppi}}\ and\ \citenamefont
  {{Rousseau}}(2006)}]{CR06}%
  \BibitemOpen
  \bibfield  {author} {\bibinfo {author} {\bibfnamefont {B.}~\bibnamefont
  {{Coppi}}}\ and\ \bibinfo {author} {\bibfnamefont {F.}~\bibnamefont
  {{Rousseau}}},\ }\href@noop {} {\bibfield  {journal} {\bibinfo  {journal}
  {\apj}\ }\textbf {\bibinfo {volume} {641}},\ \bibinfo {pages} {458} (\bibinfo
  {year} {2006})}\BibitemShut {NoStop}%
\bibitem [{\citenamefont {{Bisnovatyi-Kogan}}\ and\ \citenamefont
  {{Lovelace}}(2001)}]{Bisno01}%
  \BibitemOpen
  \bibfield  {author} {\bibinfo {author} {\bibfnamefont {G.~S.}\ \bibnamefont
  {{Bisnovatyi-Kogan}}}\ and\ \bibinfo {author} {\bibfnamefont {R.~V.~E.}\
  \bibnamefont {{Lovelace}}},\ }\href@noop {} {\bibfield  {journal} {\bibinfo
  {journal} {\nar}\ }\textbf {\bibinfo {volume} {45}},\ \bibinfo {pages} {663}
  (\bibinfo {year} {2001})}\BibitemShut {NoStop}%
\bibitem [{\citenamefont {{Piran}}(1999)}]{ReviewPiran}%
  \BibitemOpen
  \bibfield  {author} {\bibinfo {author} {\bibfnamefont {T.}~\bibnamefont
  {{Piran}}},\ }\href@noop {} {\bibfield  {journal} {\bibinfo  {journal}
  {\physrep}\ }\textbf {\bibinfo {volume} {314}},\ \bibinfo {pages} {575}
  (\bibinfo {year} {1999})}\BibitemShut {NoStop}%
\bibitem [{\citenamefont {{Krolik}}(1999)}]{LibroAGN}%
  \BibitemOpen
  \bibfield  {author} {\bibinfo {author} {\bibfnamefont {J.~H.}\ \bibnamefont
  {{Krolik}}},\ }\href@noop {} {\emph {\bibinfo {title} {{Active galactic
  nuclei: from the central black hole to the galactic environment}}}}\
  (\bibinfo  {publisher} {Princeton Univ. Press},\ \bibinfo {year}
  {1999})\BibitemShut {NoStop}%
\bibitem [{\citenamefont {{Lynden-Bell}}(1996)}]{L96}%
  \BibitemOpen
  \bibfield  {author} {\bibinfo {author} {\bibfnamefont {D.}~\bibnamefont
  {{Lynden-Bell}}},\ }\href@noop {} {\bibfield  {journal} {\bibinfo  {journal}
  {\mnras}\ }\textbf {\bibinfo {volume} {279}},\ \bibinfo {pages} {389}
  (\bibinfo {year} {1996})}\BibitemShut {NoStop}%
\bibitem [{\citenamefont {Spruit}(2010)}]{Sp08}%
  \BibitemOpen
  \bibfield  {author} {\bibinfo {author} {\bibfnamefont {H.}~\bibnamefont
  {Spruit}},\ }\href@noop {} {\bibfield  {journal} {\bibinfo  {journal}
  {Lect.Notes Phys.}\ }\textbf {\bibinfo {volume} {794}},\ \bibinfo {pages}
  {233} (\bibinfo {year} {2010})}\BibitemShut {NoStop}%
\bibitem [{\citenamefont {Velikhov}(1959)}]{V59}%
  \BibitemOpen
  \bibfield  {author} {\bibinfo {author} {\bibfnamefont {E.}~\bibnamefont
  {Velikhov}},\ }\href@noop {} {\bibfield  {journal} {\bibinfo  {journal} {Sov.
  Phys. JETP}\ }\textbf {\bibinfo {volume} {36}},\ \bibinfo {pages} {995}
  (\bibinfo {year} {1959})}\BibitemShut {NoStop}%
\bibitem [{\citenamefont {{Chandrasekhar}}(1960)}]{Cha60}%
  \BibitemOpen
  \bibfield  {author} {\bibinfo {author} {\bibfnamefont {S.}~\bibnamefont
  {{Chandrasekhar}}},\ }\href {\doibase 10.1073/pnas.46.2.253} {\bibfield
  {journal} {\bibinfo  {journal} {Proceedings of the National Academy of
  Science}\ }\textbf {\bibinfo {volume} {46}},\ \bibinfo {pages} {253}
  (\bibinfo {year} {1960})}\BibitemShut {NoStop}%
\bibitem [{\citenamefont {{Balbus}}\ and\ \citenamefont
  {{Hawley}}(1991)}]{Bal91}%
  \BibitemOpen
  \bibfield  {author} {\bibinfo {author} {\bibfnamefont {S.~A.}\ \bibnamefont
  {{Balbus}}}\ and\ \bibinfo {author} {\bibfnamefont {J.~F.}\ \bibnamefont
  {{Hawley}}},\ }\href {\doibase 10.1086/170270} {\bibfield  {journal}
  {\bibinfo  {journal} {\apj}\ }\textbf {\bibinfo {volume} {376}},\ \bibinfo
  {pages} {214} (\bibinfo {year} {1991})}\BibitemShut {NoStop}%
\bibitem [{\citenamefont {{Balbus}}\ and\ \citenamefont
  {{Hawley}}(1998)}]{B98}%
  \BibitemOpen
  \bibfield  {author} {\bibinfo {author} {\bibfnamefont {S.~A.}\ \bibnamefont
  {{Balbus}}}\ and\ \bibinfo {author} {\bibfnamefont {J.~F.}\ \bibnamefont
  {{Hawley}}},\ }\href@noop {} {\bibfield  {journal} {\bibinfo  {journal}
  {\rmp}\ }\textbf {\bibinfo {volume} {70}},\ \bibinfo {pages} {1} (\bibinfo
  {year} {1998})}\BibitemShut {NoStop}%
\bibitem [{\citenamefont {Shakura}(1973)}]{S73}%
  \BibitemOpen
  \bibfield  {author} {\bibinfo {author} {\bibfnamefont {N.~I.}\ \bibnamefont
  {Shakura}},\ }\href@noop {} {\bibfield  {journal} {\bibinfo  {journal} {\SA}\
  }\textbf {\bibinfo {volume} {16}},\ \bibinfo {pages} {756} (\bibinfo {year}
  {1973})}\BibitemShut {NoStop}%
\bibitem [{\citenamefont {{Shakura}}\ and\ \citenamefont
  {{Sunyaev}}(1973)}]{Shakura:1973p110}%
  \BibitemOpen
  \bibfield  {author} {\bibinfo {author} {\bibfnamefont {N.~I.}\ \bibnamefont
  {{Shakura}}}\ and\ \bibinfo {author} {\bibfnamefont {R.~A.}\ \bibnamefont
  {{Sunyaev}}},\ }\href@noop {} {\bibfield  {journal} {\bibinfo  {journal}
  {\aap}\ }\textbf {\bibinfo {volume} {24}},\ \bibinfo {pages} {337} (\bibinfo
  {year} {1973})}\BibitemShut {NoStop}%
\bibitem [{\citenamefont {Coppi}(1994)}]{C94}%
  \BibitemOpen
  \bibfield  {author} {\bibinfo {author} {\bibfnamefont {B.}~\bibnamefont
  {Coppi}},\ }\href@noop {} {\bibfield  {journal} {\bibinfo  {journal} {Plasma
  Phys. Contrl. Fus.}\ }\textbf {\bibinfo {volume} {36}},\ \bibinfo {pages}
  {B107} (\bibinfo {year} {1994})}\BibitemShut {NoStop}%
\bibitem [{\citenamefont {{Ogilvie}}(1997)}]{Ogilvie97}%
  \BibitemOpen
  \bibfield  {author} {\bibinfo {author} {\bibfnamefont {G.~I.}\ \bibnamefont
  {{Ogilvie}}},\ }\href@noop {} {\bibfield  {journal} {\bibinfo  {journal}
  {\mnras}\ }\textbf {\bibinfo {volume} {288}},\ \bibinfo {pages} {63}
  (\bibinfo {year} {1997})}\BibitemShut {NoStop}%
\bibitem [{\citenamefont {{Coppi}}(2007)}]{cproceeding}%
  \BibitemOpen
  \bibfield  {author} {\bibinfo {author} {\bibfnamefont {B.}~\bibnamefont
  {{Coppi}}},\ }\href@noop {} {\bibfield  {journal} {\bibinfo  {journal}
  {Proceedings of the 2007 Conference on Plasma Physics of the European
  Physical Society}\ } (\bibinfo {year} {2007})}\BibitemShut {NoStop}%
\bibitem [{\citenamefont {{Lattanzi}}\ and\ \citenamefont
  {{Montani}}(2010)}]{lattanzimontani10}%
  \BibitemOpen
  \bibfield  {author} {\bibinfo {author} {\bibfnamefont {M.}~\bibnamefont
  {{Lattanzi}}}\ and\ \bibinfo {author} {\bibfnamefont {G.}~\bibnamefont
  {{Montani}}},\ }\href@noop {} {\bibfield  {journal} {\bibinfo  {journal}
  {\epl}\ }\textbf {\bibinfo {volume} {89}},\ \bibinfo {pages} {39001}
  (\bibinfo {year} {2010})}\BibitemShut {NoStop}%
\bibitem [{\citenamefont {{Coppi}}(2009)}]{CoppiHighEnergy}%
  \BibitemOpen
  \bibfield  {author} {\bibinfo {author} {\bibfnamefont {B.}~\bibnamefont
  {{Coppi}}},\ }\href@noop {} {\bibfield  {journal} {\bibinfo  {journal}
  {\ppcf}\ }\textbf {\bibinfo {volume} {51}},\ \bibinfo {pages} {124007}
  (\bibinfo {year} {2009})}\BibitemShut {NoStop}%
\bibitem [{\citenamefont {Montani}\ and\ \citenamefont
  {Carlevaro}(2010)}]{MontaniCarlevaro10}%
  \BibitemOpen
  \bibfield  {author} {\bibinfo {author} {\bibfnamefont {G.}~\bibnamefont
  {Montani}}\ and\ \bibinfo {author} {\bibfnamefont {N.}~\bibnamefont
  {Carlevaro}},\ }\href@noop {} {\bibfield  {journal} {\bibinfo  {journal}
  {\pre}\ }\textbf {\bibinfo {volume} {82}},\ \bibinfo {pages} {025402}
  (\bibinfo {year} {2010})}\BibitemShut {NoStop}%
\bibitem [{\citenamefont {{Coppi}}(2008)}]{Coppi:2008p98}%
  \BibitemOpen
  \bibfield  {author} {\bibinfo {author} {\bibfnamefont {B.}~\bibnamefont
  {{Coppi}}},\ }\href@noop {} {\bibfield  {journal} {\bibinfo  {journal}
  {\epl}\ }\textbf {\bibinfo {volume} {82}},\ \bibinfo {pages} {19001}
  (\bibinfo {year} {2008})}\BibitemShut {NoStop}%
\bibitem [{\citenamefont {{Bisnovatyi-Kogan}}\ and\ \citenamefont
  {{Lovelace}}(2000)}]{2000ApJ...529..978B}%
  \BibitemOpen
  \bibfield  {author} {\bibinfo {author} {\bibfnamefont {G.~S.}\ \bibnamefont
  {{Bisnovatyi-Kogan}}}\ and\ \bibinfo {author} {\bibfnamefont {R.~V.~E.}\
  \bibnamefont {{Lovelace}}},\ }\href {\doibase 10.1086/308288} {\bibfield
  {journal} {\bibinfo  {journal} {\apj}\ }\textbf {\bibinfo {volume} {529}},\
  \bibinfo {pages} {978} (\bibinfo {year} {2000})},\ \Eprint
  {http://arxiv.org/abs/arXiv:astro-ph/9902344} {arXiv:astro-ph/9902344}
  \BibitemShut {NoStop}%
\bibitem [{\citenamefont {{Coppi}}\ and\ \citenamefont
  {{Coppi}}(2001)}]{Coppi:2001p102}%
  \BibitemOpen
  \bibfield  {author} {\bibinfo {author} {\bibfnamefont {B.}~\bibnamefont
  {{Coppi}}}\ and\ \bibinfo {author} {\bibfnamefont {P.~S.}\ \bibnamefont
  {{Coppi}}},\ }\href@noop {} {\bibfield  {journal} {\bibinfo  {journal}
  {\prl}\ }\textbf {\bibinfo {volume} {87}},\ \bibinfo {pages} {051101}
  (\bibinfo {year} {2001})}\BibitemShut {NoStop}%
\bibitem [{\citenamefont {Coppi}(1977)}]{ballooning2}%
  \BibitemOpen
  \bibfield  {author} {\bibinfo {author} {\bibfnamefont {B.}~\bibnamefont
  {Coppi}},\ }\href {\doibase 10.1103/PhysRevLett.39.939} {\bibfield  {journal}
  {\bibinfo  {journal} {\prl}\ }\textbf {\bibinfo {volume} {39}},\ \bibinfo
  {pages} {939} (\bibinfo {year} {1977})}\BibitemShut {NoStop}%
\bibitem [{\citenamefont {Dobrott}\ \emph {et~al.}(1977)\citenamefont
  {Dobrott}, \citenamefont {Nelson}, \citenamefont {Greene}, \citenamefont
  {Glasser}, \citenamefont {Chance},\ and\ \citenamefont
  {Frieman}}]{Ballooning1}%
  \BibitemOpen
  \bibfield  {author} {\bibinfo {author} {\bibfnamefont {D.}~\bibnamefont
  {Dobrott}}, \bibinfo {author} {\bibfnamefont {D.~B.}\ \bibnamefont {Nelson}},
  \bibinfo {author} {\bibfnamefont {J.~M.}\ \bibnamefont {Greene}}, \bibinfo
  {author} {\bibfnamefont {A.~H.}\ \bibnamefont {Glasser}}, \bibinfo {author}
  {\bibfnamefont {M.~S.}\ \bibnamefont {Chance}}, \ and\ \bibinfo {author}
  {\bibfnamefont {E.~A.}\ \bibnamefont {Frieman}},\ }\href {\doibase
  10.1103/PhysRevLett.39.943} {\bibfield  {journal} {\bibinfo  {journal}
  {\prl}\ }\textbf {\bibinfo {volume} {39}},\ \bibinfo {pages} {943} (\bibinfo
  {year} {1977})}\BibitemShut {NoStop}%
\bibitem [{\citenamefont {Montani}\ and\ \citenamefont
  {Benini}(2011)}]{2010GReGr.tmp..112M}%
  \BibitemOpen
  \bibfield  {author} {\bibinfo {author} {\bibfnamefont {G.}~\bibnamefont
  {Montani}}\ and\ \bibinfo {author} {\bibfnamefont {R.}~\bibnamefont
  {Benini}},\ }\href@noop {} {\bibfield  {journal} {\bibinfo  {journal} {\grg}\
  }\textbf {\bibinfo {volume} {43}},\ \bibinfo {pages} {1121} (\bibinfo {year}
  {2011})},\ \bibinfo {note} {10.1007/s10714-010-1038-9}\BibitemShut {NoStop}%
\bibitem [{\citenamefont {{Montani}}\ and\ \citenamefont
  {{Carlevaro}}(2010)}]{2010PhRvE..82b5402M}%
  \BibitemOpen
  \bibfield  {author} {\bibinfo {author} {\bibfnamefont {G.}~\bibnamefont
  {{Montani}}}\ and\ \bibinfo {author} {\bibfnamefont {N.}~\bibnamefont
  {{Carlevaro}}},\ }\href {\doibase 10.1103/PhysRevE.82.025402} {\bibfield
  {journal} {\bibinfo  {journal} {\pre}\ }\textbf {\bibinfo {volume} {82}},\
  \bibinfo {pages} {025402} (\bibinfo {year} {2010})},\ \Eprint
  {http://arxiv.org/abs/1007.5401} {arXiv:1007.5401 [astro-ph.HE]} \BibitemShut
  {NoStop}%
\bibitem [{\citenamefont {{Ferraro}}(1937)}]{F37}%
  \BibitemOpen
  \bibfield  {author} {\bibinfo {author} {\bibfnamefont {V.~C.~A.}\
  \bibnamefont {{Ferraro}}},\ }\href@noop {} {\bibfield  {journal} {\bibinfo
  {journal} {\mnras}\ }\textbf {\bibinfo {volume} {97}},\ \bibinfo {pages}
  {458} (\bibinfo {year} {1937})}\BibitemShut {NoStop}%
\end{thebibliography}%
\bibliographystyle{apsrev4-1}

\end{document}